\documentclass[a4paper,fleqn,usenatbib]{mnras}

\usepackage{graphicx}	
\usepackage{amsmath,amsfonts,amssymb,bm}
\usepackage{makeidx}
\usepackage{epstopdf} 
\usepackage{booktabs,caption,fixltx2e}
\usepackage[flushleft]{threeparttable}
\usepackage{hyperref}
\usepackage{multirow}
\usepackage{rotating}
\usepackage{color}
\usepackage[T1]{fontenc}
\usepackage{ae,aecompl}
\usepackage[plain]{fancyref}
\usepackage[nolist]{acronym}

\graphicspath{{figs/}}

\def \BE{\begin{equation}}
\def \EE{\end{equation}}	
\def \BC{\begin{center}}
\def \EC{\end{center}}
\def \BEA{\begin{eqnarray}}
\def \EEA{\end{eqnarray}}

\newcommand{\mr}[1]{\mathrm{#1}}

\newcommand{\dc}{\Delta}
\newcommand{\ysz}{Y_\text{\ac{sze}}}

\newcommand{\mpc}{\text{Mpc}}

\newcommand{\msun}{\text{M}_\odot}

\newcommand{\kpc}{\text{kpc}}

\newcommand{\me}{m_\text{e}}
\newcommand{\rfive}{R_\text{500c}}

\newcommand{\mfive}{M_\text{500c}}

\newcommand{\nele}{n_\text{e}}

\begin{document}

\title[SZE Observables, Pressure Profiles and Center Offsets]{SZE Observables, Pressure Profiles and Center Offsets in {\it Magneticum} Simulation Galaxy Clusters}
\author[Gupta et al.]{N.~Gupta\thanks{Nikhel.Gupta@Physik.LMU.de}$^{1,2,3}$,
A.~Saro$^{1,3}$,
J.J.~Mohr$^{1,2,3}$,
K.~Dolag$^{1,3,4}$
and
J.~Liu$^{1,5}$
\\
$^{1}$ Faculty of Physics, Ludwig-Maximilians-Universit\"at, Scheinerstr. 1, 81679 Munich, Germany \\
$^{2}$ Max Planck Institute for Extraterrestrial Physics, Giessenbachstrasse, 85748 Garching Germany\\
$^{3}$ Excellence Cluster Universe, Boltzmannstr.\ 2, 85748 Garching, Germany \\
$^{4}$ Max Planck Institute for Astrophysics, Karl-Schwarzschild-Str. 1, 85741 Garching Germany\\
$^{5}$ Bosch Research and Technology Center North America, CA 94304, United States\\
}
\maketitle

\begin{acronym}
\acrodef{mcmc}[MCMC]{Monte Carlo Markov Chain}
\acrodef{sze}[SZE]{Sunyaev-Zel'dovich effect}
\acrodef{mps}[MPS]{{\it Magneticum Pathfinder Simulation}}
\acrodef{cmb}[CMB]{cosmic microwave background}
\acrodef{icm}[ICM]{intracluster medium}
\acrodef{hse}[HSE]{hydrostatic equilibrium}
\acrodef{spt}[SPT]{South Pole Telescope}
\acrodef{tcmb}[$T_\mr{CMB}$]{CMB temperature}
\acrodef{sph}[SPH]{smoothed-particle hydrodynamics}
\acrodef{act}[ACT]{Atacama Cosmology Telescope}
\acrodef{sams}[SAMs]{Semi Analytic Models}
\end{acronym}

\begin{abstract}
We present a detailed study of the galaxy cluster thermal \ac{sze} signal $Y$ and pressure profiles using {\it Magneticum} Pathfinder hydrodynamical simulations. 
With a sample of 50,000 galaxy clusters ($\mfive>1.4\times10^{14}\msun$) out to $z=2$, we find significant variations in the shape of the pressure profile with mass and redshift and present a new generalized NFW model that follows these trends.  We show that the thermal pressure at $R_{\rm 500c}$ accounts for only 80~percent of the pressure required to maintain hydrostatic equilibrium, and therefore even idealized hydrostatic mass estimates would be biased at the 20~percent level.  
We compare the cluster \ac{sze} signal extracted from a sphere with different virial-like radii, a virial cylinder within a narrow redshift slice and the full light cone, confirming small scatter ($\sigma_{\ln Y}\simeq 0.087$) in the sphere and showing that structure immediately surrounding clusters increases the scatter and strengthens non self-similar redshift evolution in the cylinder.  Uncorrelated large scale structure along the line of sight leads to an increase in the \ac{sze} signal and scatter that is more pronounced for low mass clusters, resulting in non self-similar trends in both mass and redshift and a mass dependent scatter that is $\sim0.16$ at low masses.  The scatter distribution is consistent with log-normal in all cases.  We present a model of the offsets between the center of the  gravitational potential and the \ac{sze} center that follows the variations with cluster mass and redshift.
\end{abstract}
\acresetall


\section{Introduction}
\label{sec:ysz-introduction}
The formation and evolution of galaxy clusters are sensitive to the expansion history of the universe and to the growth rate of structure. This makes them a promising avenue to constrain different cosmological models \citep[e.g.][]{haiman01}. 
In recent years, the \ac{sze}, the inverse Compton scattering of \ac{cmb} photons by hot electrons in galaxy clusters  \citep{sunyaev70, sunyaev72}, has emerged as a powerful tool to detect massive clusters out to high redshift. This distorts the \ac{cmb} Planckian spectrum such that, at frequencies lower than 217~GHz, we observe a decrement in \ac{cmb} flux in the direction of galaxy clusters (peaking in amplitude at 150~GHz), which enables their detection \citep[see][]{birkinshaw84, rephaeli95, carlstrom02}. 

Since \citet{staniszewski09} presented the first \ac{sze} selected clusters, ongoing surveys in microwave bands such as the South Pole Telescope (SPT), the Atacama Cosmology Telescope (ACT) and Planck have yielded hundreds to thousands of newly discovered clusters \citep[e.g.][]{vanderlinde10, sehgal11, reichardt13, hasselfield13, bleem15, planck16SZcat}. Combined with the knowledge of cluster mass from follow-up programs and from simulations, 
these cluster samples provide competitive cosmological constraints \citep{planck16-24, dehaan16}.
However, current \ac{sze} cluster cosmology is limited by our understanding of cluster selection and mass-observable scaling relations \citep[see, e.g., ][]{bocquet15,planck16-24}.  In particular, further progress requires that we develop an improved understanding of cluster pressure profiles and the expected form of mass-observable scaling relations--- including the distribution of scatter and its dependence on cluster mass, redshift and radius.
 

A crucial issue in the calibration of the $\ysz$-mass relation \citep{johnston07,george12a, du14, schrabback16} and also in understanding multiwavelength \ac{sze}, optical and X-ray scaling relations \citep{Biesiadzinski12, sehgal13, planck13-11,rozo14, rozo14a, rozo14b, saro15, saro16} is the miscentering of the observable with respect to the center of mass of the cluster.  While it has been shown that approximately 80~percent of clusters exhibit good agreement in their X-ray/\ac{sze} and brightest cluster galaxy (BCG) centers \citep{lin04b, sanderson09, stott12, song12b,saro15}, providing a strong suggestion that these trace the center of mass, the behavior of the remainder of the population-- presumably those clusters that have undergone recent major mergers-- is more complicated.  Understanding the \ac{sze} center offset distribution for all masses and redshifts is helpful especially in analyses of stacked cluster observables where ignoring the tail of the offset distribution will lead to biases.

Assuming self similarity \citep{kaiser86} in the galaxy cluster population, the \ac{sze} flux within a fixed critical overdensity radius scales with mass as $Y\propto M^{5/3}$.  However, as shown with X-ray scaling relations \citep[e.g.][]{mohr97,mohr99}, physical processes underway in galaxies and the \ac{icm} can alter these scalings, producing relations that are non-self-similar.  Interestingly, the \ac{sze} scaling relations are expected to be much less sensitive to \ac{icm} physics.  Significant attention has been focused toward using simulations to understand the impact of \ac{icm} physics on the $\ysz$-mass scaling relation \citep[e.g.][]{desilva00, white02, mccarthy03, motl05, nagai06, bonaldi07, shaw08, battaglia11,kay12}.  Some of the simulations used in these studies are based on pure gravitational physics complemented with the runs of the so-called \ac{sams} of galaxy evolution. These \ac{sams} provide a realistic description of galaxy properties, but neither do they provide direct information on the properties of \ac{icm}, nor do they properly include the dynamical effects of the baryons on structure formation, which is highly relevant for the study of environmental effects.

Over the past decade, some studies have employed large scale hydrodynamical simulations that include only non-radiative physics \citep[e.g. the Marenostrum universe,][]{gottloaeber06} or a crude description of star formation \citep[e.g. the Millennium gas project,][]{gazzola07}. Simulations with higher-- though still moderate-- spatial resolution and better treatment of cooling and star formation are typically only realized for relatively small simulation volumes \citep[e.g.][]{borgani04, kay04}. Only small volumes (typically 25 ${\rm Mpc}/h$) are so far explored at high resolution (typically 2~${\rm Kpc}/h$) and with moderate inclusion of physical processes \citep[e.g.][]{tescari09, schaye09}. More physical processes like the description of radiative cooling of the \ac{icm}, a sub-resolution prescription to follow the formation and evolution of the stellar component and the release of energy and metals from Type II and Type Ia supernovae and asymptotic giant branch (AGB) stars are included in later simulations \citep[e.g.][]{tornatore07a, tornatore07b, fabjan08}. However, accurate modeling of active galactic nucleus (AGN) feedback \citep[e.g.][]{fabjan10}, inclusion of transport processes like thermal conduction \citep[e.g.][]{dolag04a} and consistent treatment of magnetic fields \citep[e.g.][]{dolag09} have allowed researchers to extend the comparison with observations towards radio wavelengths.  Recently, the cosmo-OWLS suite of cosmological hydrodynamical simulations \citep{lebrun14}-- which includes a range of physical processes like the UV/X-ray background, cooling, star formation, supernova feedback and AGN feedback-- has been explored to study group and cluster mass-observable scaling relations \citep{lebrun16}. 

The \ac{mps}\footnote{http://www.magneticum.org/} project involves a series of hydrodynamical simulations of different cosmological volumes covering a broad range of scales \citep{dolag13}.  These simulations allow us to examine the impact of \ac{icm} physics and to do so even within large enough volumes to enable good statistics in the study of rare, high mass structures like galaxy clusters \citep{hirschmann14, saro14, dolag15, teklu15, bocquet15b, dolag16}.
In this work, we study the pressure profiles and \ac{sze} cluster scaling relations using these simulations and compare them with observational results as well as previous hydrodynamical simulation studies. We study trends in pressure profiles with cluster mass and redshift and propose extensions to the standard model. We examine the offsets between the center of the cluster gravitational potential and the \ac{sze} center. 

In section~\ref{sec:ysz-simulation}, we describe the simulations. The pressure profiles are examined in section~\ref{sec:ysz-pressure-profile}, and we show the fitting results of the $\ysz$-mass relation in section~\ref{sec:ysz-yszscaling}. In section~\ref{sec:centroid-offset} we present the offset distributions of the gravitational \& \ac{sze} centers of clusters and we summarize our findings in section~\ref{sec:conclusions}.


%
\begin{table*}
  \caption[{\it Magneticum} simulation box]{ {\it Magneticum} simulation box used in this work. Column 1: size of the box in \textrm{Mpc}. Column 2: gravitational softening length for dark matter, \ac{icm} and star particles in \textrm{kpc}. Column 3: number of particles in the box. Column 4: mass of each dark matter, \ac{icm} and star particle. Column 4: minimum halo mass selected to construct the final catalog for this study. Column 5: number of halos with $M_\textrm{500c}\geq M_\textrm{500c, min}$ in the full simulation box. Column 6: number of halos in the light cone with $M_\textrm{500c}\geq M_\textrm{500c, min}$.}
  \begin{tabular}{lcccccccccc}
Size $L_\textrm{box}$ & \multicolumn{3}{c}{softening length (kpc)} & $N_\textrm{particles}$ & \multicolumn{3}{c}{$m_\textrm{particle}$ ($\msun$)} & $M_\textrm{500c, min}$ &$N_\textrm{box}$ & $N_\textrm{lc}$ \\ 
\textrm{Mpc} &DM&gas&stars&& DM&gas&star & ($\msun$)& $z \geq 0$ & $z \geq 0$\\ \hline
$1274$ &$10$&$10$&$5$& $2 \times1526^3$ & $1.8\times10^{10}$&$3.7\times10^{9}$&$9.3\times10^{8}$ & $1.4\times10^{14}$ & $49 311$ & $1 593$\\
\hline
  \end{tabular}
\label{tab:simu-box}
\end{table*}

\section{Simulation}
\label{sec:ysz-simulation}
MPS has been carried out as a counterpart to ongoing, multi-wavelength surveys \citep{DES05,planck06,carlstrom11} and to prepare for future datasets like those from eROSITA, Euclid and LSST \citep{merloni12,laureijs11,LSST09}.  The details about the simulations will be discussed in \cite{dolag13}, but here we briefly summarize the most relevant features used in this work.
\begin{figure*} 
\includegraphics[width=\textwidth]{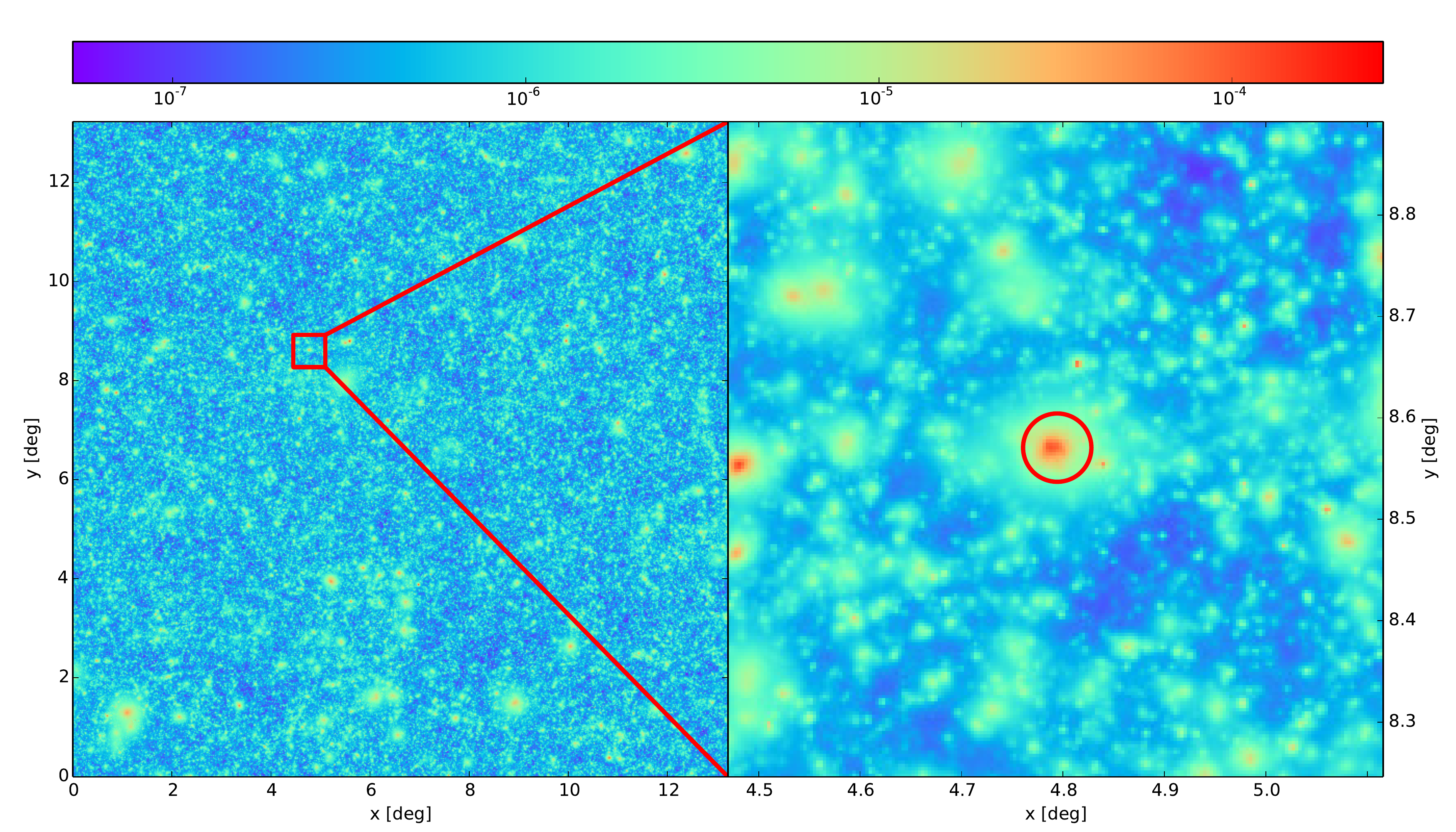}
\vskip-0.15in
\caption[Compton-y map from simulation]{Compton-$y$ map from a simulated light cone. The left panel color codes the $y(\bm\theta)$ over the whole field of view and the right panel is zoomed into one massive cluster with $\mfive=3.57\times10^{14}\msun$ at redshift $0.67$ with $\rfive$ marked with a red circle.}
\label{fig:ysz-yszmap}
\end{figure*}

\subsection{Simulation Method}
\label{sec:ysz-simulation-setup}
The simulations have been carried out with P-GADGET3, an improved version of the GADGET-2 code \citep{springel05g}. We use an entropy-conserving formulation of \ac{sph} \citep{springel02},  and a higher order kernel based on the bias corrected, sixth-order Wendland kernel \citep{dehnen12} with 295 neighbors, which together with a low-viscosity SPH scheme allows us to properly track turbulence within galaxy clusters \citep{dolag05s, donnert13, beck16}. 

Many physical processes are included in the simulations.  We include a treatment of radiative cooling computed following the same procedure as described in \citet{wiersma09}, heating by a UV background and feedback processes associated with supernovae driven galactic winds and AGN \citep{springel03, fabjan10}. We allow for isotropic thermal conduction with 1/20 of the classical Spitzer value following \citet{dolag04a}. We incorporate a detailed treatment of star formation \citep{springel03} and chemical enrichment \citep{tornatore07c}. Finally, we include passive magnetic fields based on Euler potentials \citep{dolag09}. More information about these physical processes in MPS can be found elsewhere \citep[e.g.][]{hirschmann14, bocquet15b, dolag16}.

MPS allows us to predict \ac{sze} signals from galaxy clusters with new levels of fidelity and for a large set of simulated clusters in the so-called Box1/mr (medium resolution) with $1512^{3}$ dark matter particles and the same number of gas particles in a box of $1274~\mpc$ per side. The mass of each dark matter, gas and star particle is $1.8\times10^{10}~\msun$, $3.7\times10^{9}~\msun$ and $9.3\times10^{8}~\msun$, respectively. The Plummer-equivalent softening length for gravitational forces is fixed to $10~\kpc$ in physical units from redshift $z=0$ to $z=2$. The WMAP7 flat $\Lambda$CDM cosmological parameters \citep{komatsu11} are adopted, such that the variance in the density field within 8 $h^{-1}$ Mpc $\sigma_8 = 0.809$, the Hubble constant $H_0=70.4~\rm km~s^{-1}~Mpc^{-1}$, the mean matter density $\Omega_\mathrm{m} = 0.272$, and the mean baryon density  $\Omega_\mathrm{b} = 0.0456$.

In a series of papers, the nIFTy cluster comparison project \citep[e.g.][]{elahi16, cui16, arthur16} tests eight state-of-the-art hydrodynamical codes, each equipped with their own calibrated subgrid physics and the same initial conditions are compared. They find large variations in the resulting abundance of haloes, sub-haloes and galaxies depending upon the code. They find that the dark-matter-only and non-radiative simulations do not reproduce observational results like stellar fractions and baryonic fractions in the center as well as in the in-fall region of clusters.  The P-GADGET3 code used for MPS was among those tested; we refer the reader to the nIFTY papers for further information.
 

\subsection{Compton-$y$ Map}
\label{sec:ysz-compton-y}
The \ac{sze} results in a shift of the energy distribution of the CMB photons from its blackbody spectrum. In this work, we focus on the \ac{sze} introduced by energetic electrons in the non-relativistic regime, i.e. the thermal-\ac{sze}. The amplitude of this distortion is commonly expressed in terms of \ac{tcmb}. At a given position ($\bm{\theta}$) on the sky this can be expressed as \citep{carlstrom02}
\BE \label{eq:ysz-tsz}
  \frac{\Delta\ac{tcmb}}{\ac{tcmb}}(\bm{\theta}) = y(\bm{\theta})g(x),
\EE
where $g(x)=x\coth(x/2)-4$ is a function of frequency ($\nu$) with $x\equiv h_{\rm P}\nu/k_{\rm B}\ac{tcmb}$ and $h_{\rm P}$ and $k_{\rm B}$ are the Planck and Boltzmann constants, respectively. The projected Compton-$y$ parameter is proportional to electron pressure along the line of sight
\BE \label{eq:ysz-compton-y}
  y(\bm{\theta}) = \frac{\sigma_\text{T}}{\me c^2}\int P_\text{e}(\bm\theta,l)dl,
\EE
where $\sigma_\text{T}$ is the Thompson scattering cross section, $P_\text{e}=\nele \text{k}_\text{B}T$ denotes the electron pressure and $\nele$ is the electron number density.  

The two-dimensional Compton-$y$ map is created by applying the so-called `gather approximation' with the \ac{sph} kernel \citep{monaghan85}, where all gas particles that project into the target pixel contribute to the total $y$. The gas particles are assumed to be fully ionized with mean molecular weight per free electron $\mu_\text{e}=1.14$. The details of this map-making procedure can be found in \cite{dolag05s}. To study the projection effects on the scaling relations,  we construct four light cones from randomly selected slices without rotating the simulation box. Each light cone is a stack of 27 slices extracted from the simulation box at different redshifts, such that the time interval between the slices is approximately the same ($\sim4\times10^8$~years). The field of view of each light cone is $13~\rm{deg} \times 13~\rm{deg} $  with the maximum comoving width of $\sim1228~\mpc$ at redshift $2$.  The depth of each slice along the z-direction ranges from $\sim143$ to $471~\mpc$.  We expect to have some duplicate structure at high redshift ($z>1.4$) in these light cones, but this duplication is negligible when we focus on massive clusters. Fig.~\ref{fig:ysz-yszmap} shows the resulting map from a light cone with $4096^2$~pixels.  These maps have a resolution of $\sim0.2~$~arcminute per pixel and a dynamical range in $y$ of $10^4$ from diffuse background ($5\times10^{-8}$) to massive clusters ($3\times10^{-4}$).

\subsection{Cluster Catalog}
\label{sec:ysz-cluster-cat}
Clusters are first identified using the friends-of-friends (FoF) algorithm with a linking length of 0.16 \citep[see][and references therein]{davis85}, linking only the dark matter particles. For each identified cluster halo, we then implement the spherical overdensity (SO) algorithm SUBFIND \citep{springel01, dolag09a} in parallel to compute SO masses at different overdensities like $M_{\rm 500c}$, $M_{\rm 500m}$, $M_{\rm 200c}$, $M_{\rm 200m}$ and $M_{\rm 2500c}$. Here $M_{\rm 500c}$ and $M_{\rm 500m}$, for example, describe the mass of the cluster within the region where the density is 500 times the critical and mean density of the universe, respectively. The cluster central position is recorded as the deepest gravitational potential position.  
For this analysis, the final catalog is selected to have clusters with $\mfive >1.4\times 10^{14} \msun$ and $z_{\rm max}=2$. The mean mass and redshift of this sample are $2.3\times 10^{14} \msun$ ($M_\textrm{500c, max}=2.1\times 10^{15} \msun$) and 0.31 in the full simulation box, and $2.1\times 10^{14} \msun$ ($M_\textrm{500c, max}=1.36\times 10^{15} \msun$) and 0.67 in the light cones. The details about the simulation and selected clusters are given in Table~\ref{tab:simu-box}.


\section{Pressure Profile}
\label{sec:ysz-pressure-profile}
A detailed study of the $\ysz$-mass relation of clusters requires an understanding of the pressure profile. Moreover, as mentioned already in the introduction, a matched filter cluster selection from \ac{sze} survey data can be informed by numerical simulations of clusters and their pressure profiles.  In this section, we present constraints on the \ac{icm} pressure profile and also explore potential hydrostatic mass biases by comparing the \ac{icm} pressure profile with that deduced from an \ac{hse} approximation. To analyze the pressure profiles we adopt the cluster mass definition $\mfive$.

\subsection{Pressure Profiles from the Simulations}
\label{sec:ysz-press-prof-sim}
\subsubsection{Profile Construction}
\label{sec:profile-construction}
The pressure profile of each cluster is calculated from gas particles within 30 radial bins equally spaced logarithmically between $0.1$ and $3\rfive$. We take the pressure to be the median pressure of particles in each bin, and the radial distance of the bin is the mass weighted mean radius  of the particles.  We follow the variation of the pressure about the median using the 16$^\mathrm{th}$ and 84$^\mathrm{th}$ percentiles of the distribution.  The \ac{icm} pressure profiles are constructed for $\sim 50,000$ clusters in the full simulation box (see Table~\ref{tab:simu-box}).  The mean mass and redshift of this sample are $2.3\times 10^{14} \msun$ and 0.31, respectively.  

\begin{table}\center
\caption[Constraints on GNFW Pressure Profile]{Constraints on the GNFW model parameters from fits to the pressure profiles of 50,000 simulated clusters with <$\mfive$>$=2.3\times10^{14}\msun$ and <$z$>$=0.31$.  For each parameter--- see equations~(\ref{eq:ysz-gnfw}) and (\ref{eq:ysz-p500})-- we present our results (MPS) followed by literature results from another simulation \citep{kay12} and observations \citep{planck12-10,arnaud10,mcdonald14b}.  Further discussion appears in section~\ref{sec:press-prof-fit}.}
  \begin{tabular}{lcrrrrr}
    \hline
 & \multicolumn{1}{c}{MPS} & \multicolumn{1}{c}{K12}  & \multicolumn{1}{c}{PL13} & \multicolumn{1}{c}{A10} & \multicolumn{1}{c}{McD14} \\
\hline
$P_0$ & $0.1701^{+0.0001}_{-0.0001}$& 0.33  & 0.19  & 0.21 &  $0.13^{+0.12}_{-0.05}$ \\
$c_{500}$ & $1.21^{+0.01}_{-0.01}$ & 1.97 & 1.81  & 1.18 & $2.59^{+0.74}_{-0.79}$  \\
$\gamma$  & $0.37^{+0.01}_{-0.01}$ & 0.61 & 0.31 & 0.31 & $0.26^{+0.26}_{-0.22}$  \\
$\alpha$ & $1.23^{+0.01}_{-0.01}$ & 2.04  & 1.33  & 1.05 & $1.63^{+1.01}_{-0.41}$\\
$\beta$ & $5.06^{+0.03}_{-0.03}$ & 2.99 & 4.13  & 5.19 & $3.30^{+0.86}_{-0.57}$ \\
$\alpha_\text{P}$    & $0.0105^{+0.0006}_{-0.0006}$ & $0.12$ & $0.12$ & $0.12$ & \multicolumn{1}{c}{$0.12$} \\
$c_\text{P}$   & $-0.121^{+0.002}_{-0.002}$ & \multicolumn{1}{c}{-} & \multicolumn{1}{c}{-} & \multicolumn{1}{c}{-} & \multicolumn{1}{c}{-} \\
\hline
  \end{tabular}
\label{tab:ysz-press}
\end{table}

\subsubsection{Profile Fitting}
\label{sec:press-prof-fit}

We adopt the generalized NFW model \citep[GNFW,][]{nagai07} for fitting the pressure profile. This model has been found to be a good description for the cluster \ac{icm} pressure profile in cosmological simulations \citep[e.g.][]{nagai07, kay12} and in X-ray/\ac{sze} observations of real clusters \citep[e.g.][]{arnaud10, plagge10,sun11}. The pressure $P_\text{mod}(r,M,z)$ as a function of cluster mass ($M$) and redshift ($z$) is written as
\begin{align} \label{eq:ysz-gnfw}
  P_\text{mod}(r,M,z)  = & P_{500}(M,z)\nonumber \\
 & \frac{c_{500}^\gamma(1+c_{500}^\alpha)^{(\beta-\gamma)/\alpha}}{(c_{500}~x)^\gamma[1+(c_{500}~x)^\alpha]^{(\beta-\gamma)/\alpha}},
\end{align}
where the parameters $\gamma,\:\alpha,\:\mathrm{and\:}\beta$ are the central ($r\!\!\ll\!\! r_{\rm s}$), intermediate ($r\!\!\thicksim\!\! r_{\rm s}$), and outer slopes ($r\!\!\gg\!\! r_{\rm s}$). Also, here $r_{\rm s}$=$R_{500 \rm c}/c_{500}$, $x$=$r/R_{500 \rm c}$ and $c_{500}$ is the concentration. The overall pressure scale $P_{500}$, representing the pressure at $R_{500 \rm c}$, is written as
\begin{align} \label{eq:ysz-p500}
  P_{500}(M,z)  = & 1.65\times10^{-3} P_0\, E(z)^{8/3+c_\text{P}} & \nonumber \\
&\Big[\frac{\mfive}{3\times10^{14}\msun}\Big]^{2/3+\alpha_\text{P}}~\text{keV\ cm}^{-3},
\end{align}
where $P_0$ is the dimensionless normalization and $E(z)=H(z)/H_0$. The parameters $c_{\rm P}$ and $\alpha_\text{P}$ denote the departures from self-similiarity of the pressure profile scale with redshift and mass, respectively \citep[see][]{arnaud10}.  Note that our model differs slightly from that used in previous studies; specifically, the factor $c_{500}^\gamma(1+c_{500}^\alpha)^{(\beta-\gamma)/\alpha}$ ensures that at $x=R_{500 \rm c}$, $P_\text{mod}(r,M,z)$ is equivalent to $P_{500}(M,z)$, independent of the value of $c_{500}$ and the slope parameters.

We use the Markov Chain Monte Carlo (MCMC) code, {\tt emcee} \citep[a Python implementation of an affine invariant ensemble sampler;][]{mackey13} to fit the model to the data throughout this paper.
The pressure profile is constrained by the sum of the log likelihood of the individual clusters $j$
\BE
  \label{eq:ysz-lik-pressure}
  \log \mathcal{L} = -{1\over2}\sum_{i,j} \frac{\ln[P_\text{sim}(i,j)/P_\text{mod}(r_i, M_j,z_j)]^2}{\left(\sigma ^2_\text{lnP,SPH}(i,j) + \sigma ^2_\text{lnP}(i)\right)},
\EE
where $P_\text{sim}(i,j)$ is the median pressure for cluster $j$ in different radial bins $i$ (between $0.1-3R_{500 \rm c}$) from the simulation and $P_\text{mod}(r_i, M_j,z_j)$ is the corresponding value from the model. $\sigma_\text{lnP,SPH}(i,j)$ is the log-normal particle to particle scatter determined for each cluster $j$ as half the difference between the 16$^\mathrm{th}$ and 84$^\mathrm{th}$ percentile pressure divided by the square root of the number of particles in each radial bin $i$.
The additional scatter term $\sigma_\text{lnP}(i)$ is the characteristic intrinsic logarithmic cluster to cluster scatter, which we derive iteratively from the full cluster sample. First, we calculate the cluster to cluster variation in the median pressure profile in each radial bin and adopt it while determining the best fit model.  Then we extract the cluster to cluster intrinsic scatter in each radial bin with respect to the best fit model and use that updated information to determine the best fit model again. We iterate until the resulting intrinsic scatter profile converges. 

\begin{figure}
\vskip-0.1in
\hskip-0.01\textwidth \includegraphics[width=3.4 in]{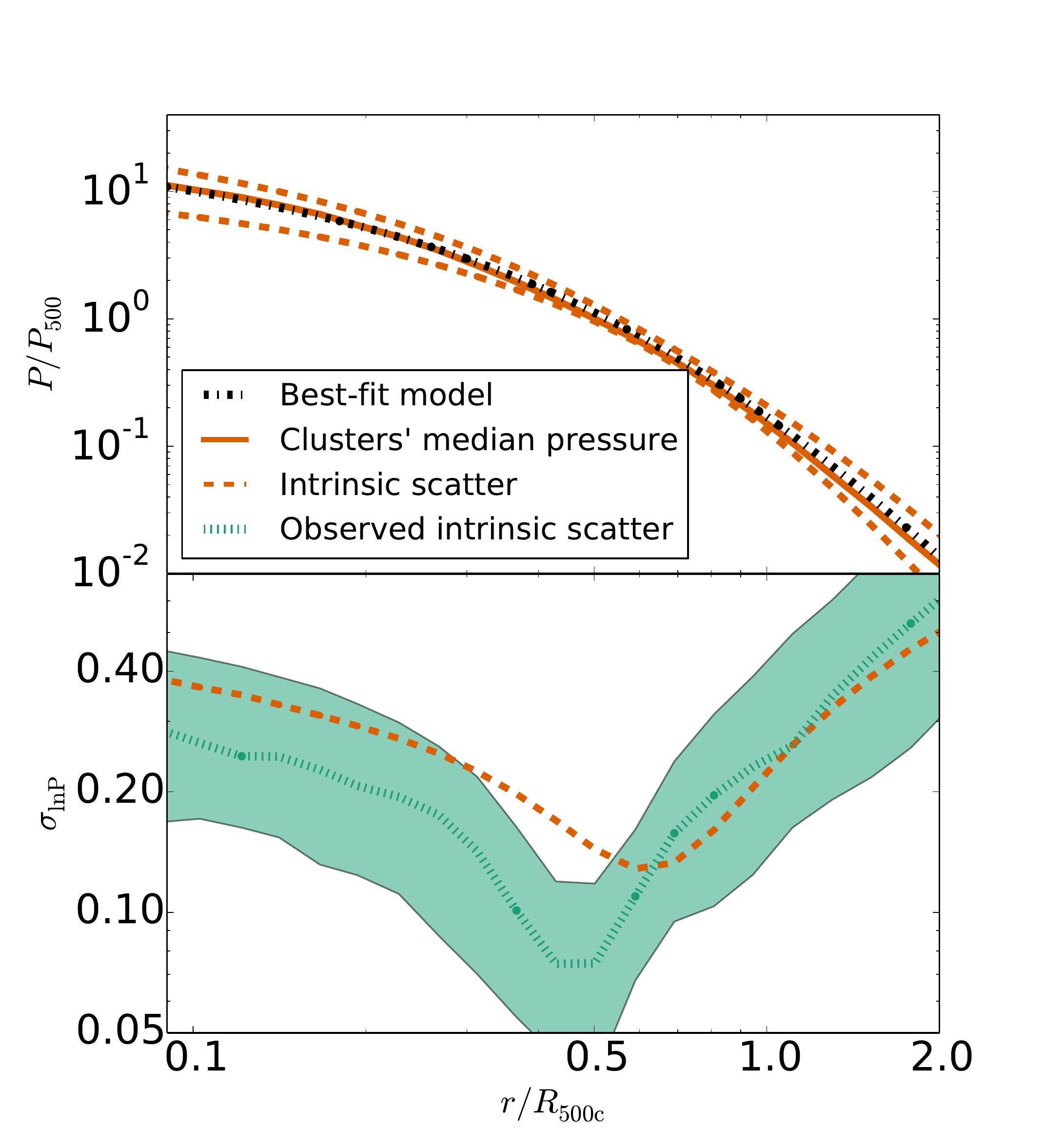}
\vskip-0.15in
\caption[The individual pressure profiles]{The best-fit pressure profile model (black dot-dashed line) derived from 50,000 clusters in the full simulation box with <$\mfive$>$=2.3\times10^{14}\msun$ and <$z$>$=0.31$. The solid red line shows the median pressure for all clusters, and the dashed red lines mark the intrinsic cluster to cluster scatter $\sigma_\text{lnP}$ about the best fit model.  This scatter also appears in the lower panel, where for comparison we show the intrinsic scatter of 31 clusters from \citet{arnaud10} (dotted green line).  The filled green region is the 1-$\sigma$ bootstrapped uncertainty on the observed intrinsic scatter.}
\label{fig:ysz-pressure-individual}
\end{figure}

The mean pressure profile from the simulation is shown in Fig.~\ref{fig:ysz-pressure-individual}. In the upper panel, the solid red line marks the median pressure profile of all clusters where the pressure for individual clusters in a radial bin is the median pressure of the number of particles in that bin.  The dashed red lines show the variation of the pressure profiles from cluster to cluster. Because the variation around the median pressure is much smaller than the cluster to cluster variation in pressure, we do not show it in the plot.  The model parameters are reported in \Fref{tab:ysz-press} as the most likely values with 68~percent confidence intervals.  Note that $P_0$ and the uncertainty for these reference studies is re-normalized to take into account the factor of $c_{500}^\gamma(1+c_{500}^\alpha)^{(\beta-\gamma)/\alpha}$ as in equation~(\ref{eq:ysz-p500}).  In calculating the rescaled uncertainties, we do not apply corrections for the degeneracies among the slope parameters.

The lower panel of Fig.~\ref{fig:ysz-pressure-individual} shows the derived intrinsic scatter as a function of radius. For comparison, we also present the observed intrinsic scatter (along with 1-$\sigma$ error bars derived by bootstrapping) in the individual pressure profiles of 31 galaxy clusters presented in table~C.1 by \citet{arnaud10}. The scatter is large in both the central and the outer part of the cluster and reaches a minimum at about $0.5R_{500}$.  We expect that this is because of the variable AGN activity in the cluster cores \citep[see, e.g.,][]{gupta16} and because of the cluster mergers and infalling field population in the outer regions. The cluster to cluster variation in the pressure profiles dominates over the cluster specific scatter within each radial bin, and therefore the latter does not impact the best fit model.

\begin{figure*}
\includegraphics[width= \textwidth]{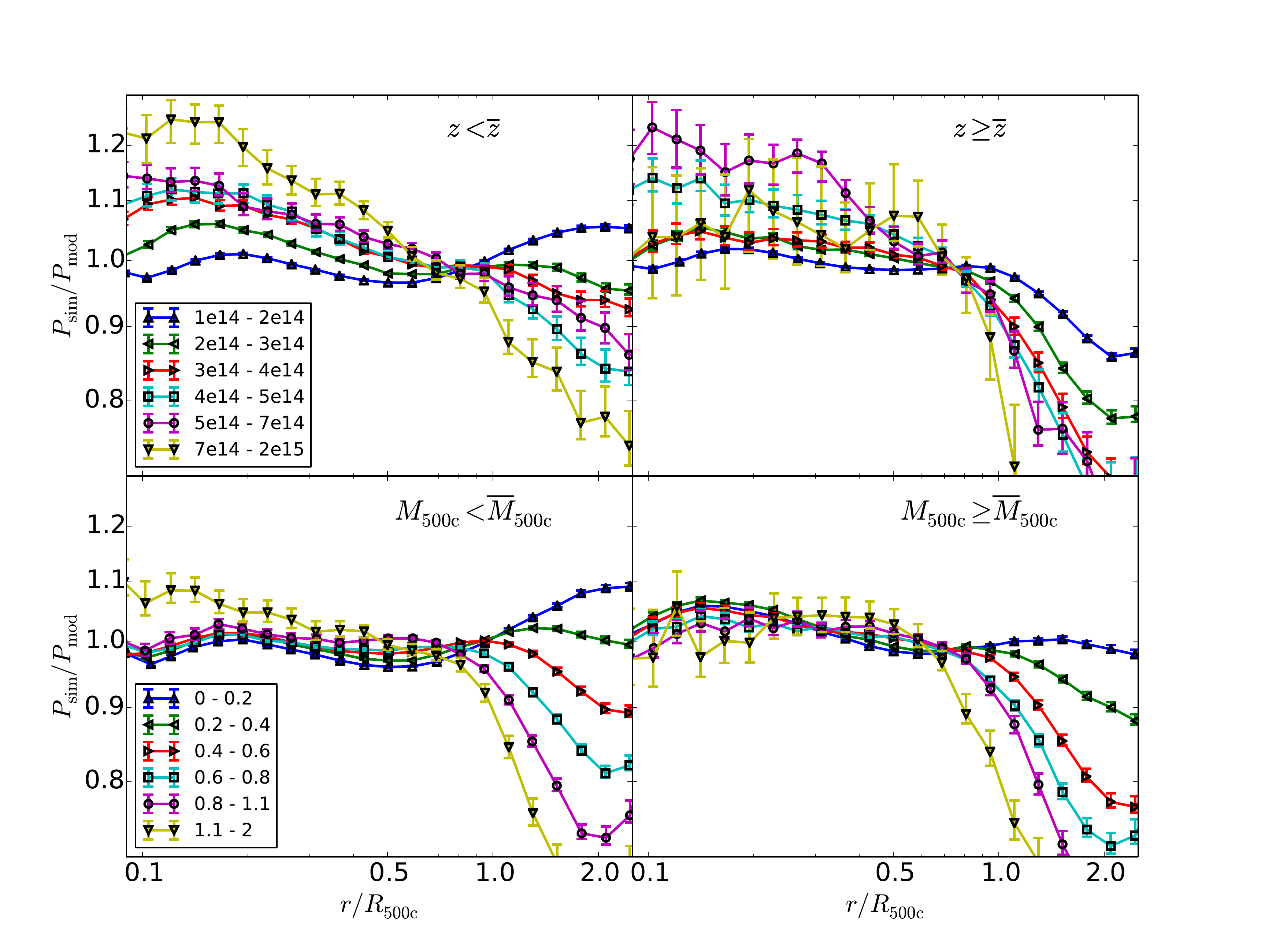}
\vskip-0.1in
\caption[The pressure profile trends]{Pressure comparison in bins of mass (above) and redshift (below) of clusters with respect to the best fit model from the full sample.  When probing for trends in mass (redshift) we subdivide the sample around the median value of redshift (mass) that is equivalent to 0.3 ($2.3\times 10^{14} \rm M_{\odot}$).  The data points mark the median of the ratio between pressure from simulations and best fit model in the radial bins and the error bars are 1-$\sigma$ uncertainties. There are clear indications of trends in mass and redshift which are more significant in the inner and outer regions of clusters, indicating that there is no universal pressure profile.}
\label{fig:pressure-trends}
\end{figure*}

\subsubsection{Variations of Profile Shape With Mass and Redshift}
\label{sec:press-mass-z-dependence}
As a next step we examine whether the \ac{icm} pressure profiles exhibit systematic shape variations with mass and redshift.  To probe for mass dependent trends we explore the behavior of the pressure profiles within six mass bins within each of two redshift ranges: $z < \overline{z}$ \& $z \geq \overline{z}$, where $\bar{z}$ is the median redshift of the whole sample.  To probe for redshift dependent trends we study the behavior of the pressure profiles within six redshift bins for each of two mass ranges:  $M_{500 \rm c}<  \overline{M}_{500 \rm c}$ \& $M_{500 \rm c}\geq \overline{M}_{500 \rm c}$, where $\overline{M}_{500 \rm c}$ is the median mass of the whole sample.  The clusters are selected from the full simulation box and divided into various bins so that there are at least 200 clusters in each bin. For the bins where the number of clusters is very large, we select a random subset of 200 clusters.

We calculate the ratios between the pressure from simulations and the best fit model (as described in section~\ref{sec:press-prof-fit}) within each of the bins. Fig.~\ref{fig:pressure-trends} shows the trends in pressure as a function of mass (above) and redshift (below).  The points represent the median of the ratios in each radial bin, and the error bars indicate the uncertainties in the median that are calculated as the standard deviation in each bin divided by the square root of the number of clusters in the bin.

These plots show clear mass and redshift trends in the shape of the \ac{icm} pressure profile, especially in the inner and outer parts of clusters. These differences are what drive the larger intrinsic scatter measured in these radial regions of clusters (as shown in the lower panel of Fig.~\ref{fig:ysz-pressure-individual} and discussed in section~\ref{sec:press-prof-fit}).  These variations are due to the trends in AGN activity and its impact on the core and trends within the infall regions with redshift and mass.  AGN provide feedback in central cluster regions, impacting the pressure profile in a mass dependent manner.  In the outskirts, deviations from the model also vary with mass and redshift, in agreement with a simulation study by \citet{battaglia12a}, where they examined a detailed dependency of  pressure profiles on cluster radius, mass and redshift.  At different redshifts, the inner regions of clusters show better self similarity as compared to cluster outskirts where the deviations increase with increasing redshift because of larger mass accretion rate at early times \citep{shi14}. Following \citet{lau15}, where they investigate the self-similarity of the diffuse X-ray emitting \ac{icm} in the outskirts of galaxy clusters, we normalize our pressure profiles using mean density ($\rho_{ \rm m}$) of universe instead of critical density ($\rho_{\rm c}$) to see if that has any impact.  We also find better self similarity in the outer pressure profiles at different redshifts, when mean density is used. However, the inner profiles become less self similar as compared to the scenario where critical density is used. This behavior may be an indication that the outer gas profiles are dependent on the late time mass accretion, which is governed by the mean density of the universe, whereas the inner profiles are dependent on the gravitational potential, which is set when the universe was still matter dominated and stays roughly constant afterwards \citep{lau15}. 

These highly significant variations of median pressure profile with mass and redshift indicate that there is no universal pressure profile.  Indeed, the pressure profile model we adopt in our analysis of the full sample is insufficient to follow the effects of complex, location dependent physics within the cluster population. We explore a few possible extensions to this simple GNFW model and present the best performing one for these MPS simulations in appendix~\ref{sec:extended_press_model}.   
We also probe for trends with redshift and mass in the evolution of the intrinsic scatter in the pressure profile deduced from the best fit model, but we find no evidence for such trends.

\begin{figure}
\hskip-0.01\textwidth \includegraphics[width=3.5 in]{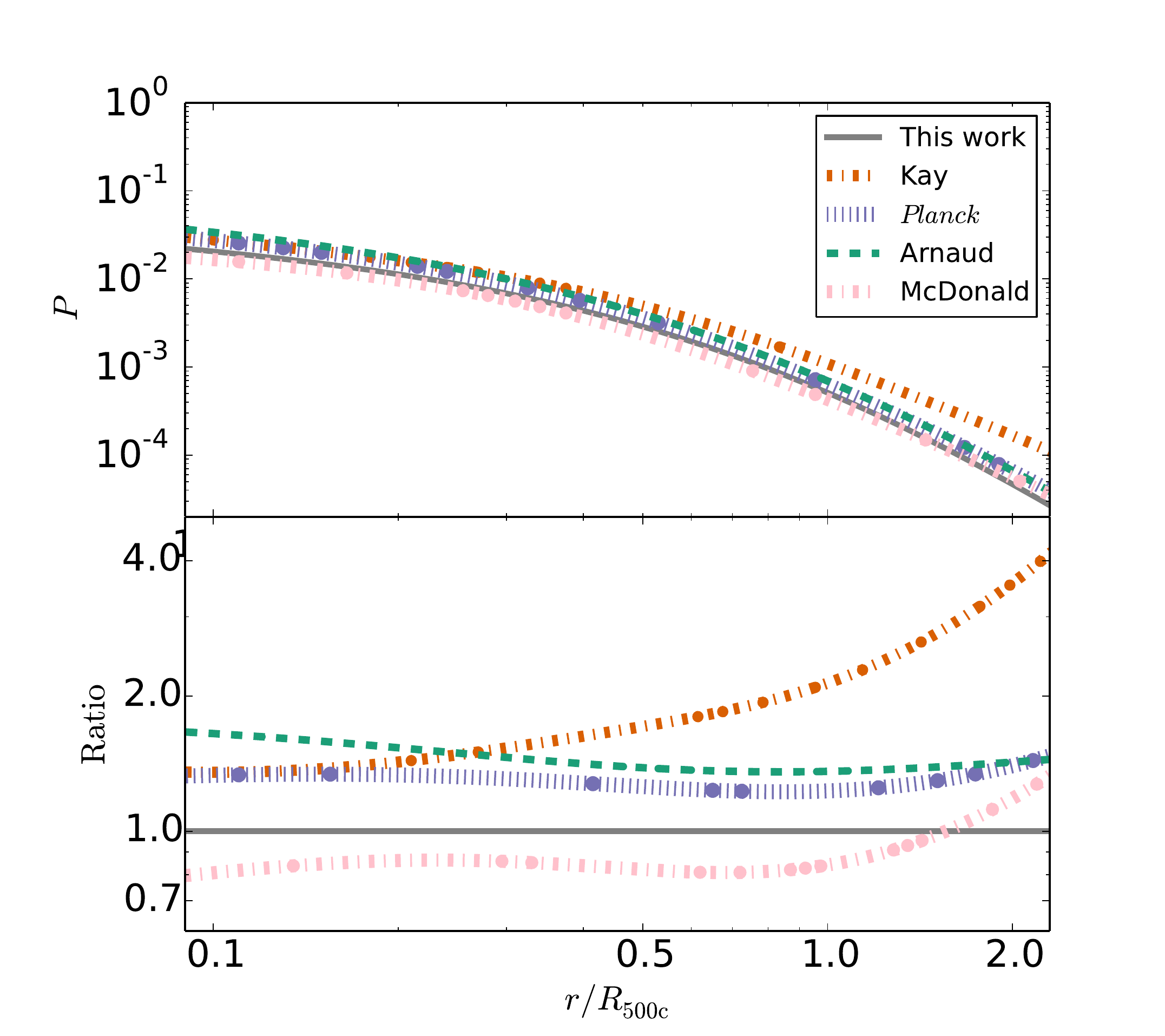}
\vskip-0.15in
\caption[The pressure profile comparison]{The comparison of the pressure profiles from the {\it Magneticum} simulations with the profiles from \cite{kay12}, \cite{arnaud10}, \cite{planck12-10} and \cite{mcdonald14b}. These profiles are constructed for a cluster mass of $\mfive = 5\times10^{14}\msun$ at $z=0$. The best fit values are stated in \Fref{tab:ysz-press}. The lower panel shows the ratio between the pressure profiles with respect to the profile obtained in this work.}
\label{fig:pressure_comparison}
\end{figure}

\subsubsection{Comparison With Previous Studies}
\label{sec:pressure-comparison}
We compare our results with previous simulation \citep{kay12} and observational studies \citep{planck12-10,arnaud10,mcdonald14b}, taking care to re-normalize to our functional form to allow for a direct comparison.  The parameter constraints from our sample and the best fit from their analysis are shown in \Fref{tab:ysz-press} and Fig.~\ref{fig:pressure_comparison}.

On the simulation side, \citet{kay12} analyzed the {\it Millennium Gas Simulations} complemented with SAMs of galaxy formation and found that the cluster \ac{icm} pressure profile can be well described by the GNFW model. They consider a feedback-only (FO) model \citep{short09} for analyzing the \ac{sze} properties of clusters. Their sample is split into high mass ($\mfive>6.8\times10^{14}\msun$) and low mass ($1.37\times10^{14}\msun \leq \mfive \leq 6.8\times10^{14}\msun$) clusters at $z=0$ and 1.  

On the observational side, pressure profiles have been reconstructed \citep{arnaud10} from X-ray observations of the REXCESS cluster sample \citep{boehringer07} at low redshift for $r<R_{500 \rm c}$ and from numerical simulations for $r>R_{500 \rm c}$.  This allows us to compare our simulated pressure profiles to the observed profiles within the inner cluster region. The best fit parameters of the GNFW model from \cite{arnaud10} are also listed in \Fref{tab:ysz-press}. The Planck collaboration \citep{planck12-10} derived the pressure profiles using XMM-{\it Newton} data for 62 massive, nearby clusters (mostly at $z<0.3$) in a large radial range out to $3\times R_{500 \rm c}$, which allows us to compare the pressure profiles outside $R_{500 \rm c}$ as well. An SPT collaboration analysis of {\it Chandra} X-ray observations of 80 \ac{sze} selected clusters \citep{mcdonald14b} divides the sample into low-$z$ ($0.3<z<0.6$) and high-$z$ ($0.6<z<1.2$) clusters. Their analysis primarily constrains the $r<1.5R_{500 \rm c}$ region of the clusters.

The comparison among the best fitting pressure profiles is shown in Fig.~\ref{fig:pressure_comparison} for a cluster with $\mfive = 5\times10^{14}\msun$ at $z=0$.  The simulated profiles from \citet{kay12} are much flatter in the outer region of the cluster, which is reflected in the smaller outer slope ($\beta$) of the GNFW model. This parameter is found to be larger in observational studies as well as in our current work.  As mentioned in \cite{kay12}, they find higher thermal \ac{icm} pressure in the outskirts of the clusters due to the absence of radiative cooling. 

Overall, our simulated profiles are comparable in shape to the observed pressure profiles.  However, the observed profiles derived from XMM observations \citep{arnaud10,planck12-10} exhibit a systematically higher pressure at the 30 to 40~percent level. 
The {\it Chandra} derived results \citep{mcdonald14b} exhibit somewhat lower pressure at the 10 to 20~percent level in the radial range $0.1-1\times R_{500\rm c}$, where the pressure profile is well constrained by the data.  
The comparison with observed properties is, however, further complicated by the different mass calibrations adopted by different authors as discussed in detail in \citet{saro16}. As a result, differences emerge not only in the predicted pressure at fixed radius (in $R_{500 \rm c}$ units), but also on the scale associated to the characteristic radius.
In summary, there is still significant disagreement in the XMM and {\it Chandra} inferred pressure profiles, with the profiles from our simulations lying roughly in the middle.

\subsubsection{Tests of Self-Similar Scaling}
\label{sec:self-similar}

It is worth noting that we have freely varied two parameters to constrain deviations from self-similar scaling of the cluster pressure normalization $P_{500}(M,z)$ with cluster mass and redshift.  This has not been done in previous analyses.  We find that $c_\mathrm{P}=-0.121\pm0.002$, indicating that the pressure normalization scales as $P_{500}\propto E(z)^{2.55}$ rather than the self similar expectation of $8/3$. This means that the pressure at a fixed cluster mass is increasing slightly less rapidly with redshift than in a self-similar model.

The parameter that describes the deviation from self-similar scaling of the pressure with the cluster mass is also inconsistent with zero ($\alpha_{\rm P}=0.0105\pm0.0006$).  The expected increase in pressure at a fixed redshift with mass is marginally steeper than the expected $2/3$ within a self-similar model.  

The non-self-similar evolution in the pressure normalization with mass and redshift reflects the mix of complex physics in the simulation that affects the amount of \ac{icm} in the cluster virial region and its thermal energy.  \citet{avestruz16} studied a mass-limited sample of galaxy clusters from cosmological hydrodynamical simulations and showed that the departure of temperature profiles from self-similar scaling in the outskirts of clusters can be explained by non-thermal gas motions driven by mergers and accretion.  In our case the small departure from self-similar scaling with redshift could be caused, for example, by the ongoing feedback from star formation and AGN in the simulations.  In the case of the mass scaling of the pressure at $R_{500}$ it is clear that any increase in the \ac{icm} mass fraction with mass \citep[e.g.][]{mohr99} must be almost perfectly offset by a slightly lower temperature than expected within self-similarity.  These offsetting effects have indeed been noted in previous simulations \citep{kravtsov05,kravtsov06}.

\subsection{Effective Pressure $P_\text{eff}$ Assuming \ac{hse}}
\label{sec:ysz-hydrostatic-pressure}
When a galaxy cluster is in a relaxed state, \ac{hse} pertains and the pressure profile is simply related to the mass density profile of the cluster, which can be described by a regular Navarro-Frenk-White profile \citep[hereafter NFW;][]{navarro97}.  In \ac{hse} there is a balance between the pressure gradient and the centrally directed gravitation attraction of the cluster on the cluster gas

\begin{equation}
  \label{eq:ysz-hse}
  \frac{dP}{dr} = -\frac{GM(r)\rho_\text{ICM}(r)}{r^2},
\end{equation}
where $M(r)$ is the total mass enclosed within radius $r$ and $\rho_\text{ICM}(r)$ is the \ac{icm} density at radius $r$.  This relation has often been assumed in deriving the masses of galaxy clusters with X-ray observations, and those masses will only be accurate if in fact \ac{hse} pertains.  Here we derive the effective pressure profile and compare it to the actual thermal pressure profile in the simulated clusters.
\begin{figure}
\hskip-0.01\textwidth \includegraphics[width=3.5 in]{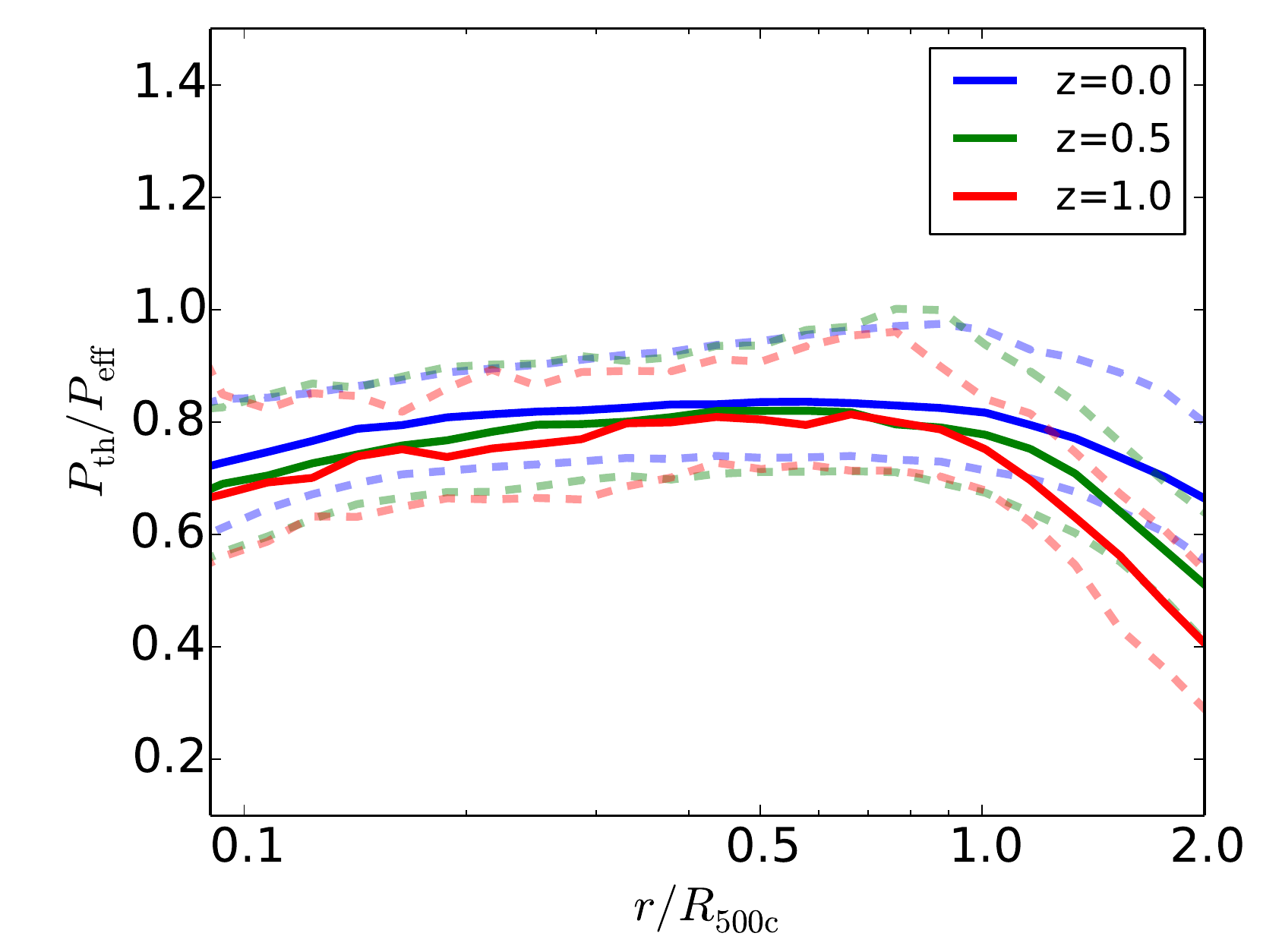}
\vskip-0.15in
\caption[Hydrostatic equilibrium pressure comparison]{Ratio of the thermal gas pressure ($P_{\rm th}$) to the \ac{hse} derived effective pressure ($P_{\rm eff}$) for all the clusters in the light cone at three different redshifts. The solid lines show the median of the ratio and the dashed lines indicate the 16$^\mathrm{th}$ and 84$^\mathrm{th}$ percentiles of the ratio, in different radial bins.}
\label{fig:ysz-pressure-hse}
\end{figure}

\subsubsection{Effective Pressure Profile Construction}
The mass of each particle, including not only dark matter and gas particles but also star and black hole particles, is summed to get the total enclosed mass as a function of radius for each simulated cluster. The gas particles are also summed separately in radial bins, providing an estimate of the gas mass that is translated into the mean gas density using the volume of the radial bins.  These ingredients, i.e the total mass and gas mass profiles, together with a boundary condition, which is the thermal gas pressure measured in the simulations at $3\times\rfive$, enable us to estimate an effective pressure profile $P_{\rm eff}(r/R_{500})$ within each cluster.  If the thermal pressure in a cluster were to match this effective pressure, the cluster would by definition be in \ac{hse}.

\subsubsection{Hydrostatic Mass Bias}
We compare the \ac{hse} derived effective pressure profiles ($P_{\rm eff}$) to the thermal gas pressure profiles ($P_{\rm th}$). Fig.~\ref{fig:ysz-pressure-hse} shows the ratio between the thermal gas pressure and the effective pressure. The median ratio from the cluster sample along with 16$^\mathrm{th}$ and 84$^\mathrm{th}$ percentile ratios (reflecting the cluster to cluster variation) is plotted in different radial bins at three different redshifts. The thermal gas pressure is always smaller than the effective pressure, implying that there must be some non-thermal pressure ($P_{\rm nth}$) support, that the cluster is still collapsing or both.  Assuming large non-thermal pressure support we can write  
\BE
P_{\rm Eff} = P_{\rm th} + P_{\rm nth}.
\EE
The median ratio of $P_{\rm th}/P_{\rm eff}$ for all clusters at a radius around $\rfive$ is $\sim80$~percent at $z=0$ and is slightly lower at $z=0.5$ and $z=1$.  At all redshifts there is a tendency for the ratio to fall at larger radii, reaching values of between 40 and 65~percent at $R_{200}$.  Because of this persistently low ratio of thermal to effective pressure, the masses obtained assuming \ac{hse} would be systematically biased low with respect to the true mass of the cluster at approximately the level of $1-P_{\rm th}/P_{\rm eff}\sim20$~percent at $R_{500 \rm c}$.  

Similar results were obtained by \citet{battaglia12}, where they estimated the \ac{hse} mass estimates from $P_{\rm th}$ and compared this mass with the true mass of clusters in hydrodynamical simulations with AGN feedback at $z=0$.  \cite{rasia12} compared weak lensing and X-ray masses of the 20 most massive simulated galaxy clusters at $z=0.25$ and noted a bias of 25-35 percent due to non-thermal pressure support and temperature in-homogeneities. \citet{biffi16} investigated the level of HSE in the intra-cluster medium of simulated galaxy clusters and found an average deviation of 10-20 percent out to the virial radius, with no evident distinction between cool-core and non-cool-core clusters. \citet{chiu12} tested the \ac{hse} assumption for cluster MS-2137 using {\it Chandra} X-ray observations combined with strong and weak lensing results from optical surveys and found a large contribution from the non-thermal pressure to the effective pressure in the cluster core, assuming a spherical model for the cluster. 

In an older work, \citet{nagai07} found that \ac{hse} estimates of the cluster mass using X-ray data can underestimate the true mass by 15~percent.  Recent comparisons of Planck collaboration hydrostatic masses to weak lensing masses over a similar redshift range indicated a tendency for the hydrostatic masses to be between 25 and 35~percent smaller, but with large uncertainties \citep{vonderlinden14, hoekstra15, planck16-24}.  Within the SPT collaboration, a comparison of hydrostatically calibrated masses to those derived from galaxy cluster velocity dispersions or from a calibration using the cluster mass function and external cosmological information also indicate that the hydrostatic masses are smaller at between 25~percent and 45~percent, respectively \citep{bocquet15}.  

In a recent work, \citet{shi15} looked into 65 clusters in a set of high-resolution cosmological hydrodynamical simulations \citep{nelson14} and found $21\pm5$~percent bias between the mass obtained assuming \ac{hse} and the true mass from simulations. Contrary to our findings, they also found a decline in $1-P_{\rm th}/P_{\rm eff}$ towards the center of the cluster which is possibly due to the non-radiative nature of simulations used in their work. AGN feedback injected in MPS would push the gas from the center that decreases the thermal component and might also add to gas motions leading to higher non-thermal pressure towards the center. However, our results are consistent towards larger radii, where $P_{\rm th}/P_{\rm eff}$ decreases as it would take a longer time to thermalize the non-thermal motions in cluster outskirts. Our results are also consistent with the fact that the non-thermal pressure support increases with redshift, perhaps due to a larger accretion rate in cluster outskirts at early times.

Several mechanisms have been proposed to understand the origin of non-thermal pressure in galaxy clusters. For instance, it has been shown that non-thermal pressure support originates from sub-sonic turbulent motions of the \ac{icm} \citep{evrard90, rasia04, dolag05, rasia06, nagai07}. \citet{fang09} and \citet{lau09}, investigating the same sample of clusters simulated by \citet{nagai07}, found that the coherent rotation of gas plays a significant role providing additional support against gravity; on the other hand, \citet{lau09} claimed that random gas motion and gas rotations have a negligible role in driving the departure from \ac{hse}.  Generally, the amount of energy in these bulk motions is of the order of 20-30~percent within the virial radius \citep{battaglia10, burn10}. Cosmic rays in clusters can also contribute to the non-thermal pressure support. Generally speaking, the contribution of cosmic rays is estimated to be less than 30 percent of the thermal pressure in the cluster core \citep{ensslin97, pfrommer08, sijacki08}.  Such a study of the origin of non-thermal pressure is beyond the scope of this paper, but we plan to study these mechanisms in future work.


\section{SZE Observable-Mass Relation} 
\label{sec:ysz-yszscaling}
In this section we present the \ac{sze} observable-mass scaling relations from our simulations and compare them to observational results and to previous studies of simulations. We analyze the $Y$-mass relation for: (1) the spherically enclosed $Y_{\rm sph}$ using different mass and virial radius definitions, (2) the cylindrical signal $Y_{\rm cyl}$ that captures the projection effects within a redshift slice of width $\sim$400~Myr (see section~\ref{sec:ysz-compton-y}), and (3) $Y_{\rm lc}$ from the projected light cones that include structure over the full redshift range extending to $z=2$.  
By comparing these three relations we hope to be able to understand the impact of correlated or nearby structures as well as uncorrelated structures randomly superposed along the line of sight.

We adopt a power law relationship between $\ysz$ and mass of the form
\BE
  \label{eq:ysz-self-similar}
  Y_\mr{sph, \dc} = 10^A  \Big[\frac{M_\dc}{3\times10^{14}\msun}
  \Big]^B \Big[\frac{E(z)}{E(0.6)}\Big]^C
  \text{Mpc}^2,
\EE
where $\Delta$ defines the overdensity used for the construction of the scaling relation. $A$ and $B$ are the fitting parameters for the normalization and mass slope of the relation, while the $C$ parameter describes the redshift evolution.  The self-similar expectations are $B=5/3$ and $C=2/3$, where this is valid for $C$ only in the case that $\Delta$ is defined with respect to the critical density.  Scaling relations of this form have been commonly used in simulations \cite[e.g.][]{kay12, battaglia12} and observations \cite[e.g.][]{planck11-10, planck13-20}, in a slightly different form. For instance, \citet{kay12} do not normalize the redshift evolution term by the factor of $E(0.6)$ as they only have clusters at $z=0$ or $z=1$, and in the Planck analysis \citep{planck11-10} they measure the intrinsic \ac{sze} signal by taking into account the angular diameter distance ($D_{\rm A}$) dependence on the observed signal. In addition to the power law scaling relation in mass and redshift listed in equation~(\ref{eq:ysz-self-similar}), we adopt a redshift and mass independent log-normal scatter $\sigma_{\ln\text{Y}}$ that is varied along with the other parameters.  

\begin{table*}
  \caption[Ysz-mass relations in spherical case]{The parameter constraints for the $Y_{\rm sph, \dc}-M_{\Delta}$ scaling relations. The {\it Planck} result is converted from Table~6 in \citet{planck11-10} to our form of the scaling relation, and the Kay result is from the feedback-only model at $z=0$ in Table~3 in \citet{kay12} where the normalization is tuned to be consistent with our relation.}
  \begin{tabular}{lrrrrr}
    \hline
    $M_\Delta$ & \multicolumn{1}{c}{$A$} & \multicolumn{1}{c}{$B$} & \multicolumn{1}{c}{$C$} & \multicolumn{1}{c}{$\sigma_{\ln Y}$} \\ 
    \hline
    $M_\text{200m}$  & -4.962$\pm$0.001 &  1.661$\pm$0.003 &   1.221$\pm$0.009 & 0.1225$\pm 0.0007$ \\ 
    $M_\text{500m}$  &  -4.829$\pm$0.002 &  1.660$\pm$0.003 &   1.386$\pm$0.013 & 0.1110$\pm 0.0009$ \\ 
    $M_\text{200c}$   & -4.896$\pm$0.001 &  1.664$\pm$0.003 &   0.543$\pm$0.010 & 0.1068$\pm 0.0009$ \\  
    $M_\text{500c}$   &  -4.758$\pm$0.002 &  1.695$\pm$0.005 &   0.571$\pm$0.015 & 0.0875$\pm 0.0011$ \\  
    $M_\text{2500c}$ &  -4.456$\pm$0.018 &   1.850$\pm$0.063 &    0.892$\pm$0.157 & 0.1361$\pm 0.0088$ \\
    \hline
    Kay $M_\text{500c}$ & -4.832$\pm$0.003 & 1.69$\pm$0.02 & \multicolumn{1}{c}{2/3} & \multicolumn{1}{c}{0.099} \\
    $Planck$ $M_\text{500c}$ & -4.769$\pm$0.013 & \multicolumn{1}{c}{1.783} & \multicolumn{1}{c}{2/3} &  \multicolumn{1}{c}{-}\\
    \hline
  \end{tabular}
  \label{tab:ysz-ym-sphere}
\end{table*}

\subsection{Spherical $Y_{\rm sph, \dc}-M_{\dc}$ Relation}\label{sec:ysz-scal-spher-ysz}
We study the spherical $Y$-mass relation for different overdensities with respect to the critical density and mean density of the universe  (as mentioned in section~\ref{sec:ysz-cluster-cat}). The  \ac{sze} signal $Y_{\rm sph, \dc}$ for each cluster is calculated using the \ac{icm} pressure distribution within a sphere of the appropriate radius centered on the cluster and is compared with the model as in equation \ref{eq:ysz-self-similar}.  The MCMC results are shown in \Fref{tab:ysz-ym-sphere} and are evaluated using all clusters in the full simulation box with mass larger than $1.4\times10^{14}\msun$ as mentioned in section~\ref{sec:ysz-cluster-cat}.

We find that the normalization $10^A$ falls systematically from the lowest mass definition $M_\text{2500c}$ (smallest radius) to the highest mass definition $M_\text{200m}$ (largest radius), but the differences are rather small.  This is due to the falling radial pressure profile as discussed in section~\ref{sec:ysz-pressure-profile}.  Interestingly, similar behavior is noticed in an analytical study by \cite{shi14}, where they show that the non-thermal pressure is increasing with mass, redshift and radius. 

The mass slope parameter $B$ at the three largest radii ($\Delta=$ 200c, 500m and 200m) is consistent with the self-similar expectation.  At the smaller radius $\Delta=$ 500c the slope is slightly steeper than self-similar and at $\Delta=$ 2500c, the smallest radius we probe here, the slope is much steeper than self-similar.   This is an indication that the physical heating and cooling processes modeled in the simulations are having a larger impact on the \ac{icm} distribution in the central regions of the cluster.
 
The redshift variation parameter $C$ shows quite a range of values.  For $\Delta$ values defined with respect to critical the self similar expectation is $2/3$, and for 200c and 500c the scaling is significantly weaker than this.  For 2500c, the central most region of the cluster, the scaling is stronger than self-similar.  Here again, the suggestion is that the cooling and heating processes modeled in the simulations are affecting the redshift evolution of the cores and outskirts of clusters in different ways.  This is consistent with our profile results that suggest the shape of the pressure profile is changing with redshift.  

For cases where $\Delta$ is defined with respect to the mean density, the redshift evolution is different because the mean density scales as $\rho\propto (1+z)^3$ as opposed to the critical density scaling as $\rho_\mathrm{crit}\propto E^2(z)$.  Because $\rho(z)=\Omega_{\rm m}(z)\rho_{\rm crit}(z)$, the scaling relations built using overdensities with respect to mean density $\Delta_{\rm m}$ exhibit not only the evolution of $\rho_{\rm crit}$ seen in the scaling relations using the critical overdensity, but also the evolution of the density parameter from $\Omega_{\rm m}=0.3$ at $z=0$ to $\Omega_{\rm m}\sim1$ at higher redshift.  This generically leads to more rapid redshift evolution in the $\Delta_{\rm m}$ relations and thus  higher values of $C$.  Moreover, the physical regions corresponding to 200m or 500m are correspondingly larger than those for 200c and 500c, and thus any differences in the evolution of the cores and outskirts will lead to differences in the redshift evolution of the critical and mean relations.

The log-normal scatter $\sigma_{\ln Y}$ falls from 0.12 at $\Delta=$ 200m to 0.09 at $
\Delta=$ 500c, corresponding to a $\sim$30~percent improvement in the regularity of these clusters within 500c as compared to 200m.  However, in the very central region $\Delta=$ 2500c the scatter is 0.14, indicating that the central core region is the most varied due to the complex physical processes included in the simulation and their impact on the cluster cores. This radial dependence of the scaling relation scatter is consistent with the radial dependence of the intrinsic scatter in the pressure profile as shown in Fig.~\ref{fig:ysz-pressure-individual}. 

In \Fref{tab:ysz-ym-sphere}, we also show the scaling relation parameters from two past analyses. Because the relations used in these studies are slightly different than ours, we convert the best fit values in-accordance with the fitting relation in equation~\ref{eq:ysz-self-similar}. \citet{kay12} vary normalization $A$ and slope $B$ of the relation, but keep the $C$ parameter -- that models the redshift evolution of the relation -- fixed to the self-similar value.  Our analysis shows that self-similar redshift evolution is not a good description of our simulated cluster ensemble. They also calculate the scatter about the best fit relation.  Our log-normal scatter about the $Y_{500\rm c}-M_{\rm 500c}$ relation is similar to theirs.  The mass slope of the relation is also consistent, preferring a scaling that is slightly steeper than self similar but with a factor of four larger uncertainty.  The preferred normalization in our analysis is $\sim$17~percent higher than theirs.

We also show observational results from the Planck analysis \citep{planck11-10} where the \ac{sze} signal is measured in the direction of $\sim 1600$ clusters from the MCXC \citep[Meta-Catalogue of X-ray detected Clusters of galaxies,][]{piffaretti11} catalog. Because this was an X-ray based analysis we expect that these results should be most comparable to our spherical $Y$ scaling relation results.  The parameters $B$ and $C$ were fixed in the analysis to values that are inconsistent with the behavior we see, but the normalization is in agreement at the 1$\sigma$ level.  Comparison to future large observed cluster samples where mass and redshift trends are left free will be very interesting.

\begin{table}
\caption[Ysz-mass relations with projection.]{The parameter constraints for the $Y-\mfive$ scaling relation for spherical and projected \ac{sze} signals. $Y_\text{sph}$ represent the spherical signal, $Y_\text{cyl}$ is the cylindrical signal in redshift slices, $Y_\text{lc}$ is the signal from the whole light cone and $Y_\text{co}$ is the signal that captures the large scale structure contribution along the line of sight, which is measured using the halo mass function. $\sigma_{\ln Y}$ is derived assuming mass and redshift independence and has the same error bars $\simeq 0.002$ for all the cases.}
\begin{tabular}{ccccc}
\hline
  \multicolumn{1}{c}{Obs} & \multicolumn{1}{c}{$A$} & \multicolumn{1}{c}{$B$} & \multicolumn{1}{c}{$C$} & \multicolumn{1}{c}{$\sigma_{\ln Y}$}\\ \hline
$Y_\text{sph}$ & $-4.753\pm 0.002$ & $1.68\pm 0.01$ & $0.55\pm  0.01$ & $0.088$ \\
$Y_\text{cyl}$ & $-4.697\pm 0.002$ & $1.65\pm 0.01$ & $0.45\pm 0.02$ & $0.102$ \\
$Y_\text{lc}$ & $-4.649\pm 0.002$ & $1.55\pm 0.01$ & $0.24\pm 0.02$ & $0.159$ \\
$\left<Y_\text{lc}\right>$ & $-4.643\pm 0.002$ & $1.51\pm 0.02$ & $0.21\pm 0.02$ & - \\ 
\hline
\end{tabular}
\label{tab:ysz-mass-proj}
\end{table}

\subsection{Cylindrical $Y_{\rm cyl}-M_{\rm 500c}$ Relation}
\label{sec:ysz-from-slice}

We investigate the projected \ac{sze} signature by studying the $Y_{\rm cyl}-M_{\rm 500c}$ relation that captures the effects of surrounding structures within a redshift slice that are projected onto the cluster. We take the clusters present in all simulation light cones and compute their projected $Y_{\rm cyl}$ within the simulated redshift slice where the cluster exists.  There can be a bias in the cluster signal due to the surrounding structure within which the clusters are embedded as well as other clusters within the slice.  To focus on the cluster associated signal, we choose only those clusters for which there is no other cluster along the line of sight and where the whole $5\rfive$ region of the cluster is contained within the redshift slice. 

The result is presented in \Fref{tab:ysz-mass-proj} along with the result from the $Y_{\rm sph}$ signal from the same cluster sample. Also note that, the fitting result of $Y_\text{sph}$ in \Fref{tab:ysz-ym-sphere} is different from \Fref{tab:ysz-mass-proj}, which is because the cluster selections are different. In the first case, we select all the clusters present in the whole simulation box at all redshifts, whereas in the second case we just select the clusters present in the light cone. In low redshift slices, there are fewer clusters in light cones due to the limited field of view, so the redshift distributions of the  two cluster samples are slightly different. 

The main difference between the scaling parameters from the spherical ($Y_\text{sph}$) and cylindrical ($Y_\text{cyl}$) signals is in the normalization. The larger signal from the cylindrical volume as compared to the spherical volume is simply evidence that the cluster \ac{sze} signature extends well outside $R_{500 \rm c}$. The mean ratio between the two measurements is $Y_\text{cyl}/Y_\text{sph}=1.151$.  In addition, the scatter is approximately 10~percent larger in the cylinder case, reflecting the additional variations introduced by the variations in the nearby structure projected onto the cluster $R_{500}$ region. Finally, the redshift evolution is less steep, suggesting that there are redshift dependent changes in the contributions to the cluster \ac{sze} signal from the surrounding structures.
\begin{figure}
\vskip-0.20in
\centering {\hbox{\includegraphics[width=90mm, height = 90mm,trim = 20 0 -20 0]{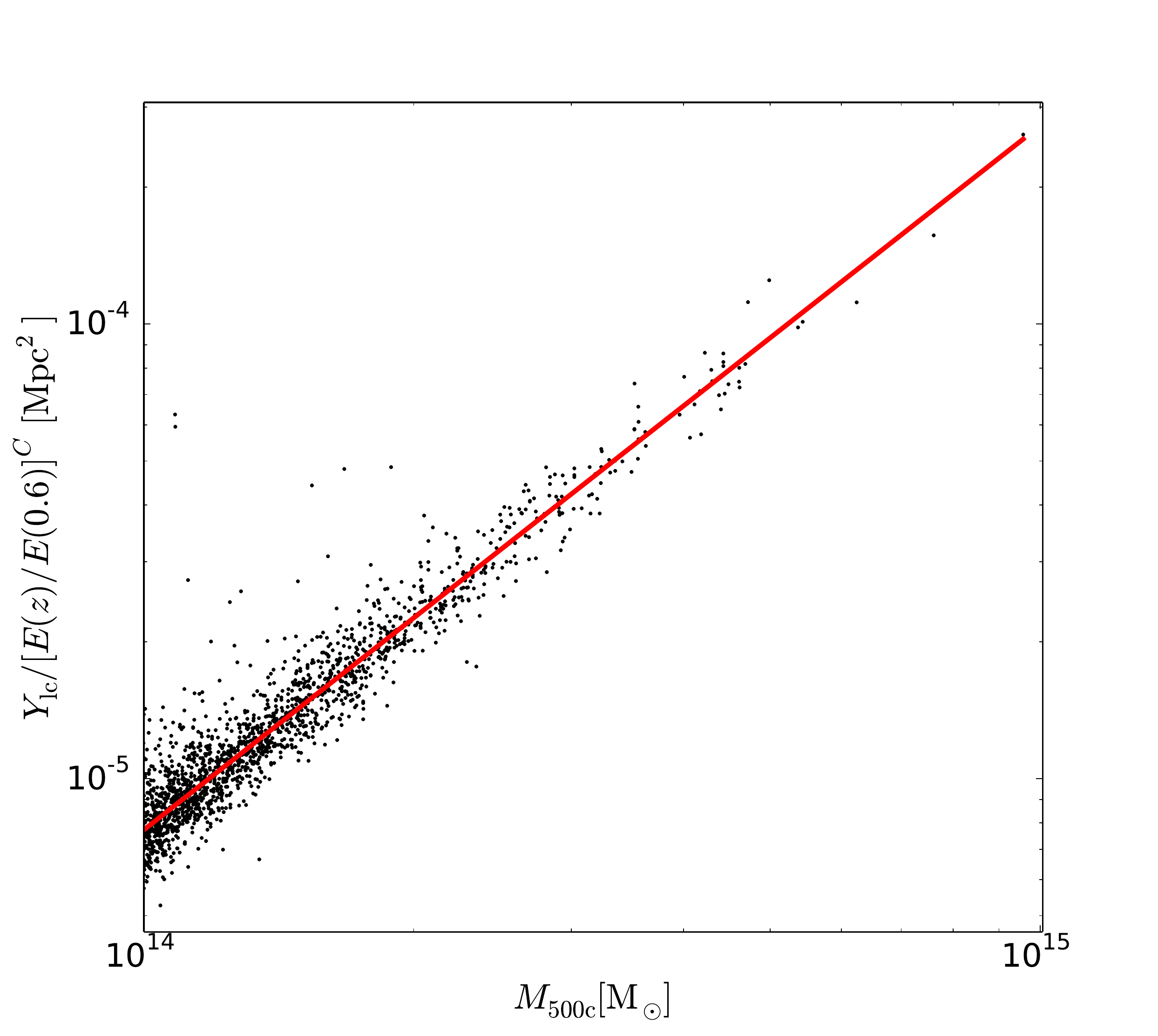}}}
\vskip-0.15in
\caption[$Y_{\rm lc}$-Mass relation]{The scaling between $Y_{\rm lc}$ and $M_{\rm 500c}$ where the black points represent clusters from all lightcones and the red solid line shows the best fit model. The scatter is clearly larger at the low mass end, whereas for $Y_\text{sph}$ and $Y_\text{cyl}$ we find no clear mass trends.  This suggests that the mass dependent scatter is caused by the varying contribution of large scale structure to the total \ac{sze} signal from a cluster, introducing a mass trend that scales as $\sigma_{\ln Y} \propto M_\text{500c}^{-0.38\pm0.05}$ (see section~\ref{sec:scatter}).}
  \label{fig:ylc-mass}
\end{figure}
\begin{figure}
\vskip-0.20in
\centering {\hbox{\includegraphics[width=90mm, height = 100mm,trim = 20 0 -20 0]{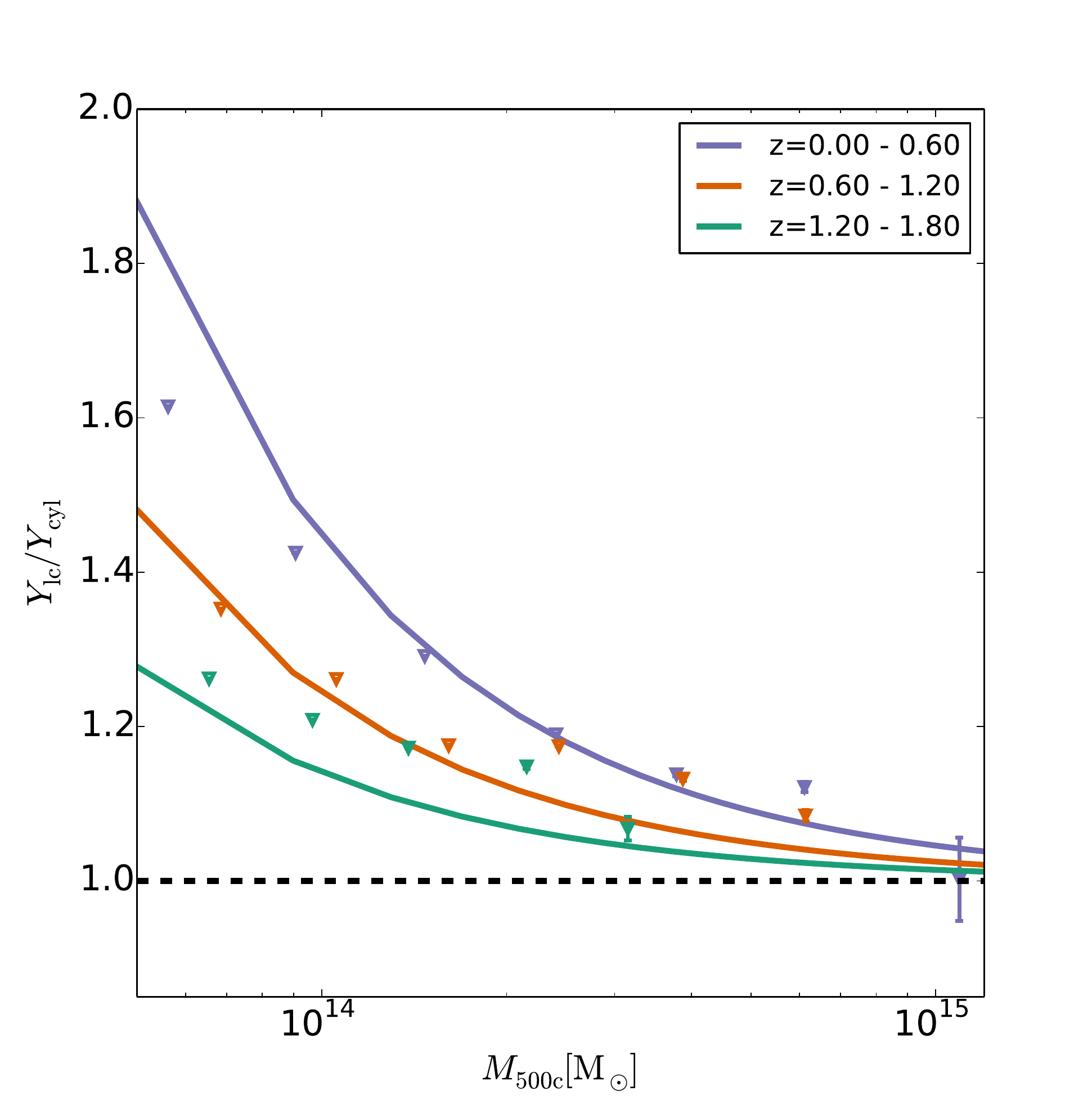}}}
\vskip-0.15in
\caption[Impacts of light cone on Y]{The ratio of the \ac{sze} signal extracted from light cones to that from cylinders within redshift shells is plotted versus mass for three different redshift bins. The points are the mean measurements from the simulated light cone, and the lines mark the expected impact from the \ac{sze} signal of uncorrelated structures along the line of sight, as described in equation~(\ref{eq:ylcest}).  Points and lines are color coded by redshift. The bias decreases with increasing cluster mass and redshift.}
  \label{fig:ysz-proj}
\end{figure}
%


\subsection{Light Cone $Y_{\rm lc}-M_{\rm 500c}$ Relation}
\label{sec:ysz-lightcone}

We also explore the scaling relation between the \ac{sze} signal extracted from light cones $Y_{\rm lc}$ and mass.  For this investigation, we take all the clusters that are completely inside the light cone boundaries. Thus, we have contributions to the \ac{sze} signal from clusters which overlap with each other along the line of sight.  \Fref{tab:ysz-mass-proj} shows the scaling relation fits derived from the $Y_{\rm lc}$ measurements, and Fig.~\ref{fig:ylc-mass} contains a plot of the relation with the redshift trend projected out. The $Y_{\rm lc}$ scaling relation deviates significantly from self-similar evolution, preferring weaker trends with mass and redshift than those we see with the spherical or cylindrical \ac{sze} observables.  In addition, in comparison to the cylindrical case, the normalization is 11~percent higher, and the scatter is a factor of 1.5 higher at 0.159.  In Fig. \ref{fig:ylc-mass}, the black data points are the clusters from all lightcones and the red solid line is the best fit model. There is a clear indication that the scatter is larger at the low mass end, behavior which was not apparent in the scaling relations involving $Y_\text{sph}$ and $Y_\text{cyl}$.  This suggests that the unassociated large scale structures along the line of sight are introducing a mass dependent scatter, a subject that we return to in the next section. 

Because $Y_{\rm lc}$ is impacted by the superposition of physically uncorrelated structure along the line of sight, one can estimate the difference between $Y_{\rm lc}$ and $Y_{\rm cyl}$ using the mean $y$ from the simulation light cone \citep[see also][]{kay12}.  We find this mean value to be $\left<y_{\rm lss}\right>=1.02\times10^{-6}$ $\rm sr^{-1}$ when averaged over four lightcones.  
Following this logic, we express the estimate for the light cone \ac{sze} signal $\left<Y_{\rm lc}\right>$ to be
\BE
\left<Y_\text{lc}\right> =Y_\text{cyl}+\left<y_{\rm lss}\right>\pi R^2_{\rm 500c},
\label{eq:ylcest}
\EE
where $R_{\rm 500c}$ is the radius of the cluster, which naturally is a function of the cluster mass and redshift.
Fig.~\ref{fig:ysz-proj} shows the ratio of the light cone to the cylindrical \ac{sze} signal $Y_{\rm lc}/Y_{\rm cyl}$ as a function of cluster mass in three redshift ranges.  Both direct measurements from simulation (points) and our simple model (line) are shown.  It is clear that the impact of the background and foreground $y$ due to projected structures is much larger on the low mass clusters.  Moreover, one can see that at a fixed mass the impact is higher at lower redshift.   

This behavior follows directly from equation~(\ref{eq:ylcest}), where it is the virial extent of clusters that determines how large a region of contaminating background is combined with the cluster \ac{sze} signal.  This virial area scales as $M_{\rm 500c}^{2/3}$, while the cylindrical signal $Y_{\rm cyl}$ scales as $M_{\rm 500c}^{5/3}$, so the biasing contribution from the background and foreground structures scales as $M_{\rm 500c}^{-1}$, becoming less important for more massive clusters.  Similarly, at fixed mass clusters have higher cylindrical \ac{sze} signal at higher redshift, scaling as $E^{2/3}(z)$, and smaller virial area at high redshift scaling as $E^{-4/3}(z)$, due to the higher density of the Universe; therefore, the contamination due to the mean $\left<y_\text{lss}\right>$ as a fraction of the cluster signal scales as $E^{-2}(z)$.  The observed behavior in the simulations, demonstrated in Fig.~\ref{fig:ysz-proj}, agrees qualitatively with this expectation.

As a consistency check, we test whether $\left<y_{\rm lss}\right>$ measured from the light cones is consistent with the sum of the \ac{sze} signals from the population of halos along the line of sight. Specifically, we express the mean $y$ due to large scale structure as
\begin{equation}
\left< y_{\rm lss}\right> = \iint Y_\mathrm{cyl}(M,z) D_\text{A}(z)^{-2}
\frac{dn}{dM}\frac{dV}{dzd\Omega} \Delta dM dz \label{eq:ysz-proj},
\end{equation}
where $\frac{dn}{dM}$ is the cluster mass function \citep{tinker08,eisenstein98}, $V$ is the volume, 
and $Y_\mathrm{cyl} (M,z)$ is the \ac{sze} signal contributed by each cluster (see section~\ref{sec:ysz-from-slice}).
We integrate over redshift from 0.001 to 2 and over mass from $10^{12}$ to $10^{15} \msun$ in equation~(\ref{eq:ysz-proj}).  The impact from extending the integral to larger redshift and lower masses is negligible for clusters with masses above $10^{14} \msun$.  Thus, we are able to recover the measured mean \ac{sze} signal in the simulations to within 20~percent accuracy through contributions from galaxy to cluster scale halos.

The best fit parameters for the $Y_{\rm lc}$-$M_{\rm 500c}$ and the $\left<Y_{\rm lc}\right>$-$M_{\rm 500c}$ are given in \Fref{tab:ysz-mass-proj}.  The amplitude, mass and redshift trends are similar for the two relations, although there is statistical tension at the $\sim2\sigma$ level in the mass and redshift slopes.  Note that to the extent that this bias is featureless, most \ac{cmb} scanning strategies from ground based instruments would remove the bulk of this background signal.  However, it is clear that departures from flatness in $\left< y_{\rm lss}\right>$ contribute significantly to the cluster scatter about the \ac{sze} mass--observable relation, and therefore one might well expect that the shifts we see in the amplitude and in the redshift and mass trends will be largely mirrored in the observations.

\begin{figure}
\vskip-0.15in
\includegraphics[width=0.5\textwidth]{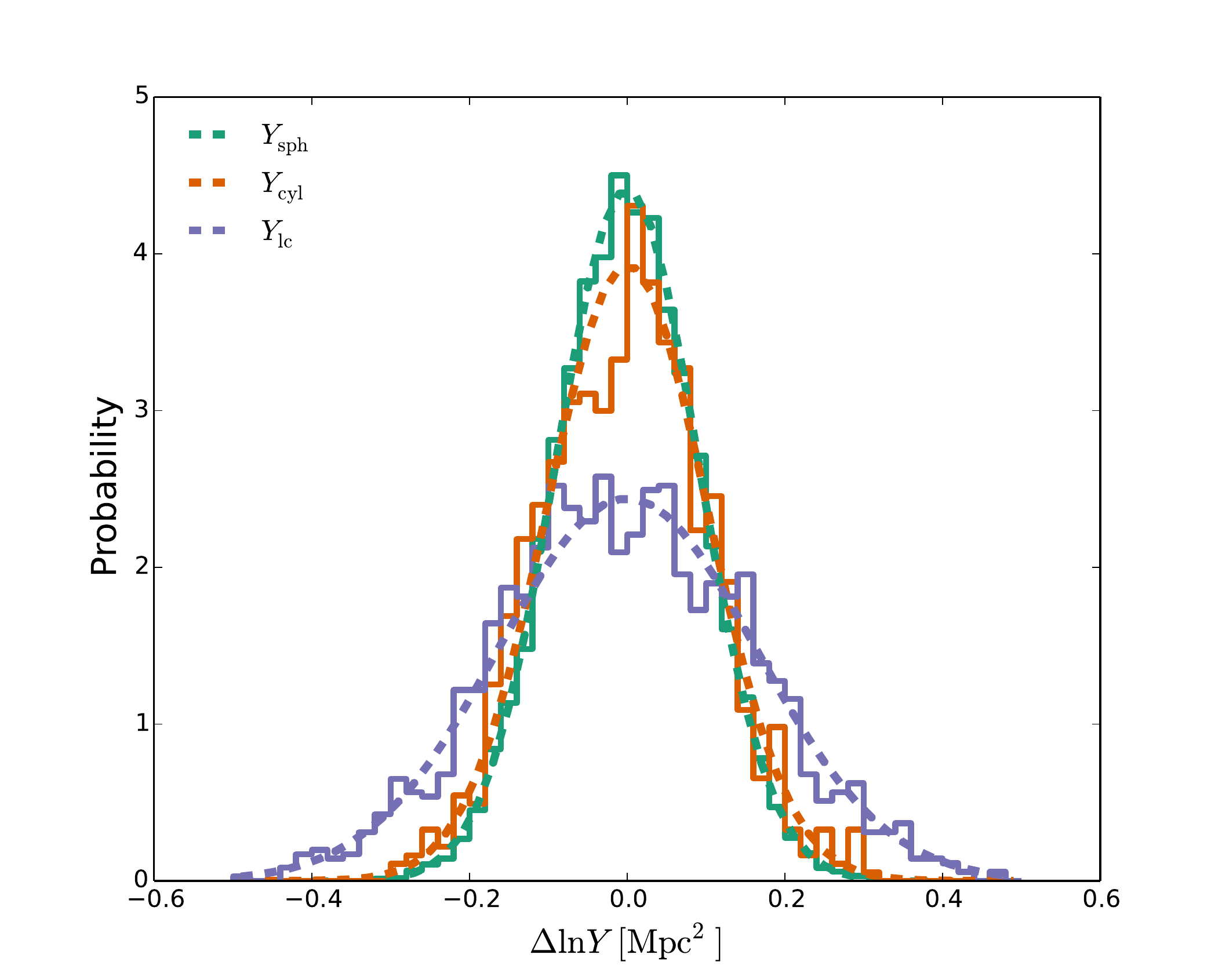}
\vskip-0.15in
\caption[Ysz residual distribution]{The distributions of scatter about the $Y_{500\rm c}-M_{500\rm c}$ scaling relations for the spherical (green), cylindrical (red) and light cone (blue) cases. The cylindrical and light cone cases show the impacts of surrounding and physically unassociated large scale structure along the line of sight, respectively. In each case a Gaussian with the same standard deviation as the scatter distribution is shown with a dashed line.  A KS test indicates that the distributions are consistent with log-normal distributions in all cases.}
\label{fig:ysz-dy-dist}
\end{figure}
\begin{figure*}
\includegraphics[width=8.8cm, height=4.6cm]{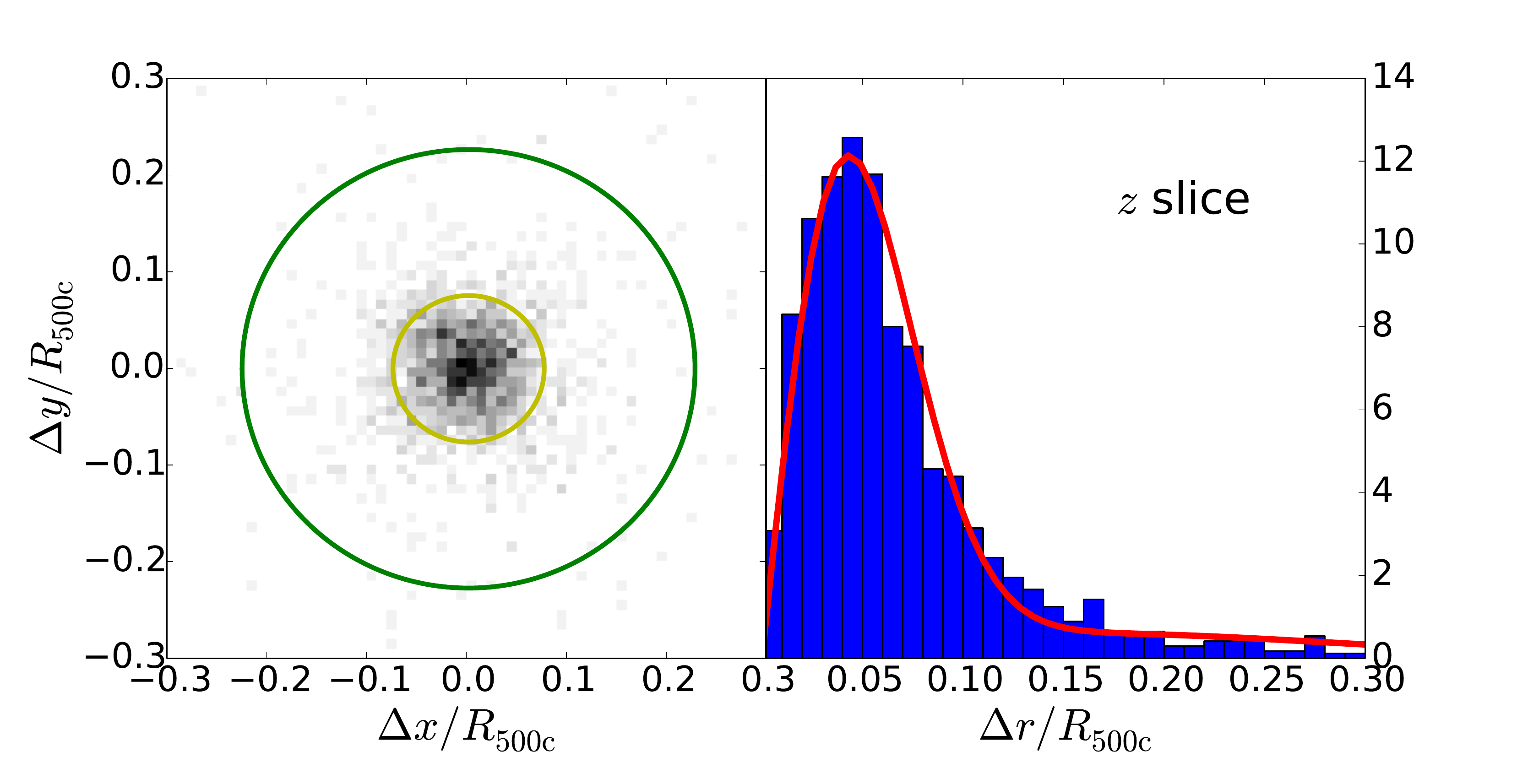}
\includegraphics[width=8.8cm, height=4.6cm]{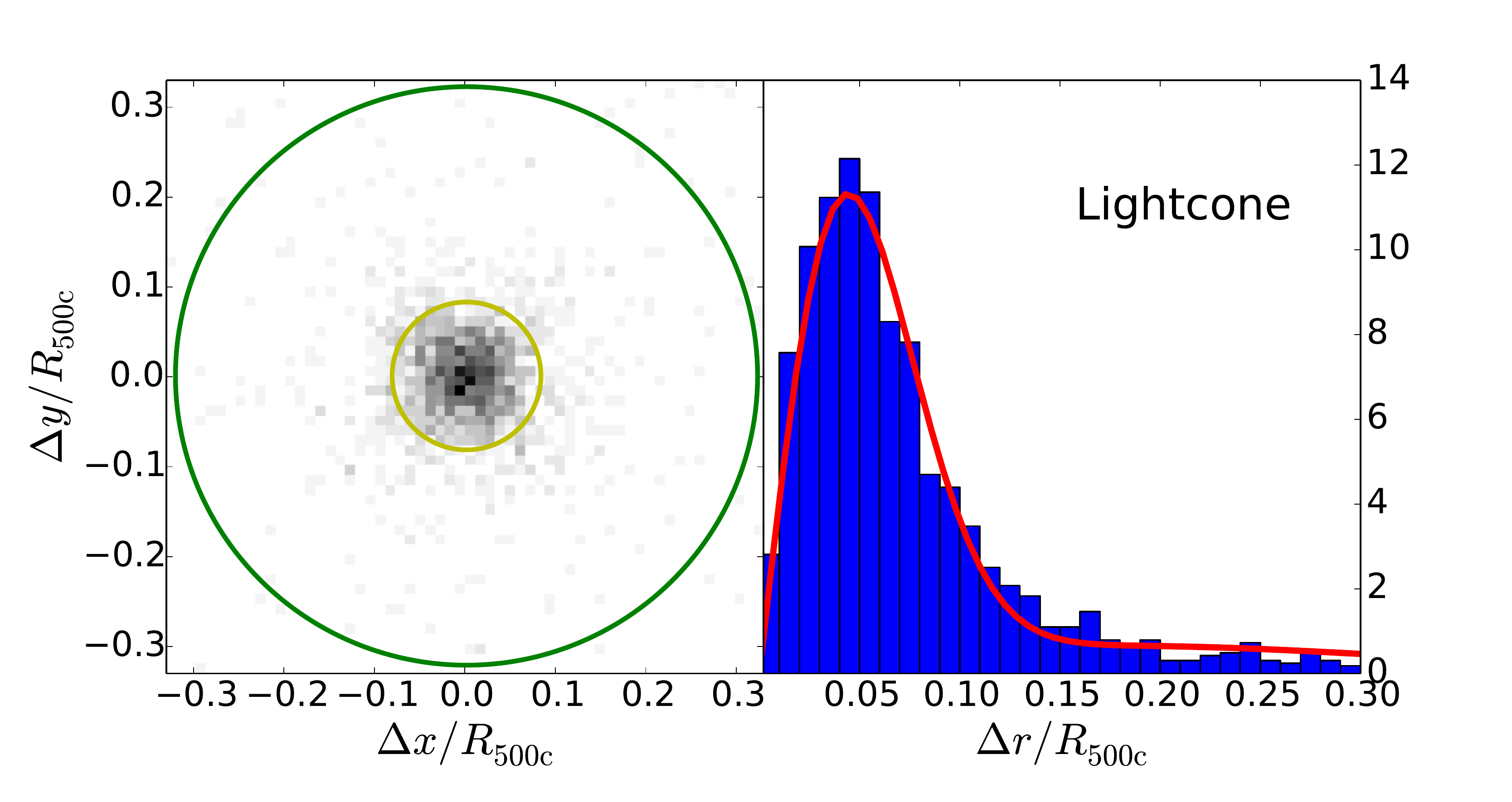}
\vskip-0.1in
\caption{The distributions of central offsets between the gravitational center and the peak of the $\ysz$ signal for clusters in redshift slices (left panels) and light cones (right panels). The yellow and green circles mark the 68$^\mathrm{th}$ and 95$^\mathrm{th}$ percentiles of the distribution, respectively. Radial offset distributions are also shown where the red line describes the double Rayleigh fit to these distributions. The fits indicate that about 80~percent of clusters populate the narrow first component with $\sigma(\Delta r/R_{500 \rm c})\sim0.045$ and the remainder populate the wider component with $\sigma\sim0.16$ (see Table~\ref{tab:central-offset}).}
  \label{fig:central-offset}
\end{figure*}

\subsection{Scatter about the $Y$-$M$ Relations}
\label{sec:scatter}

Table~\ref{tab:ysz-mass-proj} contains the log-normal constant scatter $\sigma_{\ln Y}$ about the best fit mass--observable relations at overdensity 500c, and Fig.~\ref{fig:ysz-dy-dist} shows the actual distributions of scatter.  These distributions are shown as histograms color coded for the spherical \ac{sze} signal $Y_\text{sph}$ (green), the cylindrical signal extracted from redshift shells $Y_\text{cyl}$ (red) and the light cone signal $Y_\text{lc}$ (blue).  As clear from Table~\ref{tab:ysz-mass-proj}, the scatter increases in each of these steps.  The cylindrical signal is sensitive to the asphericity of clusters and the variation in the surrounding structures in the infall regions, and the light cone includes scatter contributions from variations in the unassociated structures projected along the line of sight.  A best fit log-normal distribution is plotted (dashed lines) for each of these distributions.   A Kolmogorov-Smirnov (K-S) test provides no evidence that the scatter distributions are inconsistent with log-normal; the p-values are 0.83, 0.29 and 0.28 for the spherical, cylindrical and light cone cases, respectively.

As briefly mentioned in the previous section, Fig.~\ref{fig:ylc-mass} shows a clear trend in scatter with cluster mass.  This motivates us to fit the mass dependence of the scatter to capture the impact of the uncorrelated structures along the line of sight. We fit for a scatter of the form
\BE
\label{eq:scatter-evol}
\centering
\sigma_{\ln Y}^2= \sigma_{\ln Y_{\rm cyl}}^2  + \sigma_{\rm A}^2  \Big[\frac{M_{\rm 500c}}{3\times 10^{14}\msun}\Big]^{\sigma_{\rm B}},
\EE
along with other scaling relation parameters to all clusters in the lightcones.  This form includes a floor $\sigma_{\ln Y_{\rm cyl}}$ to the scatter, which we measure from the $Y_{\rm cyl}-M_{\rm 500c}$ relation and the quantifies the additional scatter in the light cones with a possible mass dependence.  We find $\sigma_{\rm A}=0.088 \pm 0.006$ and $\sigma_{\rm B}=-1.65 \pm 0.26$, which mean that 30 percent and 66 percent of scatter is coming from uncorrelated large scale structure along the line of sight for a cluster $M_{\rm 500c}$ of $10^{15} \rm M_{\odot}$ and $10^14 \rm M_{\odot}$, respectively.  This provides clear evidence that low mass clusters are more affected by the mean background/foreground \ac{sze} signal.  In addition to this, we also investigate trends in the scatter with redshift but find no clear evidence for that. 

Interestingly, when we probe for mass dependent scatter in the $Y_{\rm cyl}$-mass and $Y_{\rm sph}$-mass relations, we find no evidence to support it ($\sigma_{\rm B}$ consistent with zero).  This suggests that the cluster asphericity and variation in the surrounding structure have a similar fractional impact on the \ac{sze} observable for all masses and redshifts, but that the variation in the uncorrelated structures along the line of sight toward clusters has a more significant impact on the scatter of the \ac{sze} observable for low mass clusters than for high.  

In a recent study, \citet{lebrun16} presented the scaling of the spherical $Y$ signal with mass and showed a variation in the log-normal scatter by a factor of 2-3 in a mass range of $10^{13}-3\times 10^{14} \rm M_{\odot}$. They noted that such a variation in the amplitude of the scatter is caused by the inclusion of non-gravitational physics like AGN feedback that can increase or decrease the scatter depending upon the complexity of physics.  Our results here that are extracted from more massive clusters suggest a different picture.


\section{SZE Center Offsets}
\label{sec:centroid-offset}

We study the offset between the center of the gravitational potential (most bound particle) in the clusters and the centers defined to be the peak of the \ac{sze} signal.  We choose the clusters in all redshift slices and light cones with $\mfive > 1.4\times10^{14}\msun$. We further select clusters which have no overlap with other clusters along the line-of-sight and the clusters for which the whole $\rfive$ region is contained within the respective redshift slice or light cone to avoid boundary effects in the signal. 

We calculate the projected offset between the \ac{sze} center and the gravitational center in the $x$ and $y$ directions and normalize them by the cluster $R_{\rm 500c}$.   Fig.~\ref{fig:central-offset} contains a plot of the scaled offsets measured using the redshift slices (left) and the light cone (right).  Over the full sample of clusters we find the 68$^\mathrm{th}$ and the 95$^\mathrm{th}$ percentiles for the offset distributions, and these are shown in yellow and green circles, respectively.  We also show the radial offset distributions $\Delta r/R_{500 \rm c}$ (blue histograms in Fig.~\ref{fig:central-offset}).  The 68$^\mathrm{th}$ percentile value for $\Delta r/R_{\rm 500c}$ is 0.075 and 0.081 in the redshift slices and the light cone, respectively.

Following \cite{saro15}, we model the 1-D offset distributions by fitting a double Rayleigh function of the form
\BE
P(x) = 2\pi x \left( \frac{\rho_0}{2 \pi \sigma_0^2}e^{{-\frac{x^2}{2 \sigma_0^2}} } + \frac{1- \rho_0}{2 \pi \sigma_1^2}e^{{-\frac{x^2}{2 \sigma_1^2}} } \right),
\label{eq:center-off}
\EE
where $x=\Delta r/R_{\rm 500c}$, $\rho_0$ is the fraction of distribution in the first component, and $\sigma_0$ and $\sigma_1$ are the widths of the two components, respectively.  The first component indicates the relatively relaxed cluster population with small offsets, whereas the second component with larger offsets represents those systems that have undergone mergers recently.  To reduce the degeneracy between $\sigma_0$ and $\sigma_1$, we define $\sigma_1 = \Delta \sigma + \sigma_0$ and adopt $\Delta\sigma$ as the free parameter in the fit.
The best fit parameters and associated uncertainties are presented in Table~\ref{tab:central-offset} along with results from \cite{saro15}.  The reasonably good agreement between our results and the observational results suggests that the observed optical-\ac{sze} offset may well reflect the expected offsets between the \ac{sze} signal and cluster center due to the ongoing growth and evolution of clusters.
In Fig.~\ref{fig:central-offset} we show these offsets as 1-D histograms for clusters in redshift slices as well as in light cones. The red curves over the 1-D histograms are the best fit models.
\begin{figure*}
\includegraphics[width= \textwidth]{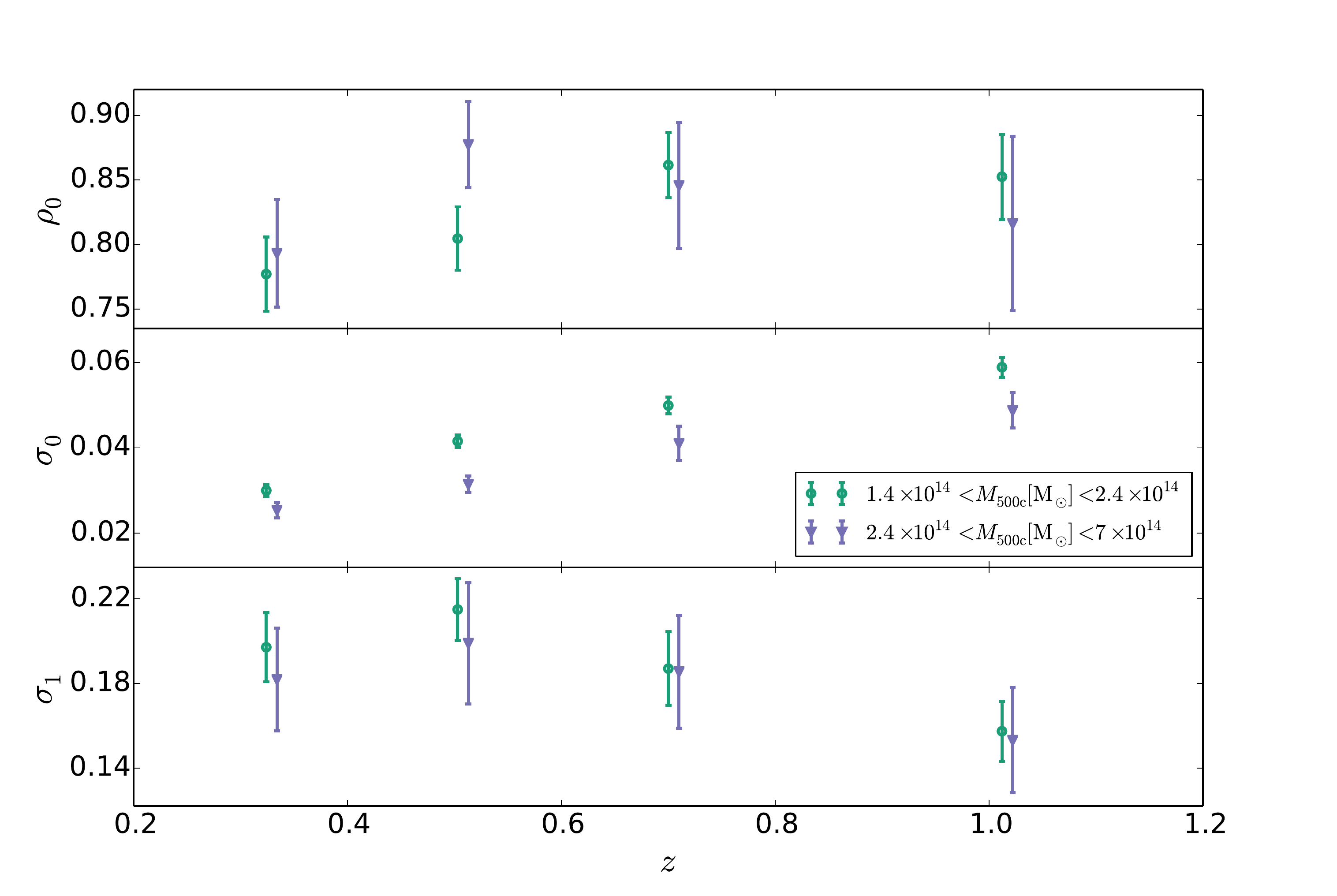}
\vskip-0.15in
\caption{The best fit parameters of the double Rayleigh function in equation (\ref{eq:center-off}) that model the radial offset distribution of the light cones for different cluster subsamples. While the fraction of large offset or disturbed clusters remains at $\sim$20~percent for all subsamples and the characteristic offset of this sample is $\sim0.18R_{\rm 500c}$, the characteristic offset for the more relaxed subset is larger for lower mass systems and grows with redshift.}
\label{fig:offset-trends}
\end{figure*}
\begin{table}
\centering
\caption{The best fit parameters of the double Rayleigh function (see equation~\ref{eq:center-off}) fit to the radial offset distribution between the gravitational potential center and the $\ysz$ peak for clusters in the redshift slices and in the light cones. In the last column we compare these numbers from an observational study by \citet{saro15}.}
\begin{tabular}{lrrr}
\hline
Parameter & \multicolumn{1}{c}{$z$-slice} & \multicolumn{1}{c}{Light cone} & \multicolumn{1}{c}{Saro15}\\ 
\hline
$\rho_0$  			&	$0.838\pm 0.013$	&	$0.802\pm 0.013$ & $0.63^{+0.15}_{-0.25}$\\
$\sigma_0$		&	$0.043\pm 0.001$	&	$0.044\pm 0.001$ & $0.07^{+0.03}_{-0.02}$\\
$\sigma_1$		&	$0.163\pm 0.006$	&	$0.184\pm 0.006$ & $0.25^{+0.07}_{-0.06}$\\
\hline
\end{tabular}
\label{tab:central-offset}
\end{table}

Next we examine the variation in the offset distributions with redshift and mass by dividing the sample into two mass and four redshift bins. We fit the double Rayleigh function (equation~\ref{eq:center-off}) to clusters in light cones for each of these bins. The best fit parameters and uncertainties are presented in Fig.~\ref{fig:offset-trends}. The fraction of clusters ($\rho_0$) in the small offset population is always consistent with the best fit value in Table~\ref{tab:central-offset} for the full sample, suggesting that the fraction of merging or disturbed clusters remains at $1-\rho_0 \sim 20$~percent, independent of the mass and redshift ranges of the sample.  Indeed, this large offset fraction is consistent with the large offset distributions seen between BCG and X-ray/\ac{sze} centers in samples of nearby X-ray selected clusters and \ac{sze} selected samples spanning a broad range of redshift \citep{lin04b,song12b}.

The characteristic offset for the disturbed population in units of $R_{\rm 500c}$ is also consistent with the best fit value from all clusters indicating that the impact of the merging clusters on the offset distribution is similar for all mass and redshift bins.  In a recent study of the X-ray morphology of samples of X-ray and \ac{sze}-selected clusters from ROSAT and SPT \citep{nurgaliev16},  no statistically significant redshift evolution in the X-ray morphology, over the range $0.3<z<1$ was found.  This is largely consistent with our result;  however, the width of the relaxed population ($\sigma_0$) shows clear trends with mass and redshift. The width is larger for low mass and more distant systems. This motivates us to include mass and redshift dependence in this parameter. We discuss a more sophisticated model in appendix~\ref{sec:revised-offset-model}. 

Finally, we also explore a possible correlation between the mean bias in pressure evaluated assuming HSE ($<P_{\rm th}/P_{\rm eff}>$ within cluster $R_{\rm 500c}$ and/or up to $3R_{\rm 500c}$) and central offsets for clusters in lightcones.  We find these quantities to be un-correlated independent of the redshift of clusters.


\section{Conclusions}
\label{sec:conclusions}
For the past few years observations of large scale structure at millimeter wavelengths have been exploited and a large sample of galaxy clusters out to high redshifts has been selected using the \ac{sze}. The counts of clusters as a function of mass and redshift provide a wealth of information about the evolution of the universe. The current cluster cosmology is, however, reliant on the calibration of mass--observable scaling relations, which enables accurate modeling of the selection and comparison to the expected mass function for each cosmology.

In this study we analyze the $Y$-mass scaling relation using the {\it Magneticum} Pathfinder hydrodynamical simulations (see section~\ref{sec:ysz-simulation-setup}). These simulations allow us to predict \ac{sze} signals from galaxy clusters and study the large scale structure projection effects for a large set of simulated clusters.

We study the thermal gas pressure profiles for high mass clusters out to high redshift using a generalized-NFW model and allowing for departures from self-similar trends in the pressure normalization $P_{500}$ (see section~\ref{sec:ysz-press-prof-sim}).  Never before such a large parameter space has been explored in studies of the cluster pressure profile.  We compare our best fit pressure profile and the measured variance among profiles as a function of radius to observed profiles and find reasonable agreement.  We study the variation in the shape of the pressure profile with cluster mass and redshift, finding large variations, with the most significant differences in the inner and outer parts of the cluster.  Thus, our analysis of these simulated clusters demonstrates that a universal pressure profile is not expected.  We present an extended version of the GNFW model that includes the trends with mass and redshift that we find (see appendix~\ref{sec:extended_press_model}).

We explore the deviation from self-similar trends with mass and redshift of the pressure normalization $P_{500}$, finding deviations from self-similarity that are statistically highly significant but nonetheless quite modest. The mass dependence of the pressure normalization is inconsistent with the self-similar value at $\sim 2$~percent and the redshift trend at $\sim$5~percent.

We study the effective cluster pressure deduced from the true cluster mass obtained from the simulations using the \ac{hse} approximation (see section~\ref{sec:ysz-hydrostatic-pressure}). The effective pressure is larger than the thermal gas pressure due to the presence of non-thermal pressure in clusters. We find $\sim 20$ percent bias between \ac{hse} derived effective pressure and the thermal gas pressure around $R_{\rm 500c}$. This implies a bias in the X-ray derived hydrostatic masses of galaxy clusters at the same level, and provides additional evidence that hydrostatic masses are not adequate for cluster cosmological studies unless this bias can be properly accounted for.

The $\ysz$-mass relation is analyzed for different mass overdensity definitions. The cluster $Y$ is extracted from the virial sphere $Y_\text{sph}$, from a cylinder within narrow redshift shells $Y_\text{cyl}$ and from a full light cone $Y_\text{lc}$ (see section~\ref{sec:ysz-yszscaling}). 
We find the $Y_\text{sph}$ scaling relation with overdensity 500c to have the least scatter ($\sigma_{\ln Y}\simeq 0.087$, see \Fref{tab:ysz-ym-sphere}), to exhibit mass trends consistent with self-similarity but redshift trends weaker than the self-similar expectation.
We analyze the impact of projection effects on the scaling relation using $Y_\text{lc}$ from the light cones,  seeing a mass and redshift dependent increase in the cluster \ac{sze} signal and a mass dependent scatter going as $\sigma_{\ln Y} \propto M_\text{500c}^{-0.38\pm0.05}$.  The \ac{sze} signal from uncorrelated structures along the line of sight can be explained through the contributions of \ac{sze} signal from halos with masses between $10^{12}$ and $10^{15}M_\odot$.  The $Y_\text{lc}$--mass relation is decidedly non-self-similar in its redshift and mass trends (see Table~\ref{tab:ysz-mass-proj}). The scatter distributions about the best fit relations are log-normal.

Finally, we analyze the central offset between the $\ysz$ signal peak and the center of the gravitational potential, modeling it as a double Rayleigh distribution (see section~\ref{sec:centroid-offset}).  We find $\sim 20$~percent of the population in the broader offset distribution, which is an indication of recent merging, while the rest exhibit small characteristic offsets of $\sim0.04R_{500c}$.  A study of trends in the Rayleigh distribution with mass and redshift shows that the relaxed population remains at $\sim$80~percent while exhibiting an increase in its characteristic offsets with decreasing mass and increasing redshift. We present a revised offset model that includes the trends with cluster mass and redshift (see appendix~\ref{sec:revised-offset-model}).

\section*{Acknowledgements}

We thank Eiichiro Komatsu, Daisuke Nagai and David Rapetti for suggestions that improved the analysis and discussion. We also thank I-Non Chiu, Adam Mantz, Raffaella Capasso, Corvin Stern, Sebastian Bocquet and Alfredo Zentano for their help in improving the final manuscript. We acknowledge the support by the Faculty Fellowship program at the Max Planck Institute for Extraterrestrial Physics, the International Max Planck School for Astrophysics,  the DFG Cluster of Excellence ``Origin and Structure of the Universe'', the Transregio program TR33 ``The Dark Universe''   and the Ludwig-Maximilians-Universit\"at.  The data processing has been carried out on the computing facilities of the Computational Center for Particle and Astrophysics (C2PAP), located at the Leibniz Supercomputer Center (LRZ).  The {\it Magneticum} Pathfinder simulations have been performed at the Leibniz-Rechenzentrum with CPU time assigned to the Project pr86re.

\appendix
\section{Additional pressure profile modeling}
\label{sec:extended_press_model}
\begin{figure*}
\hbox to \hsize{\hskip-0.4in
\includegraphics[width= 1.15\textwidth]{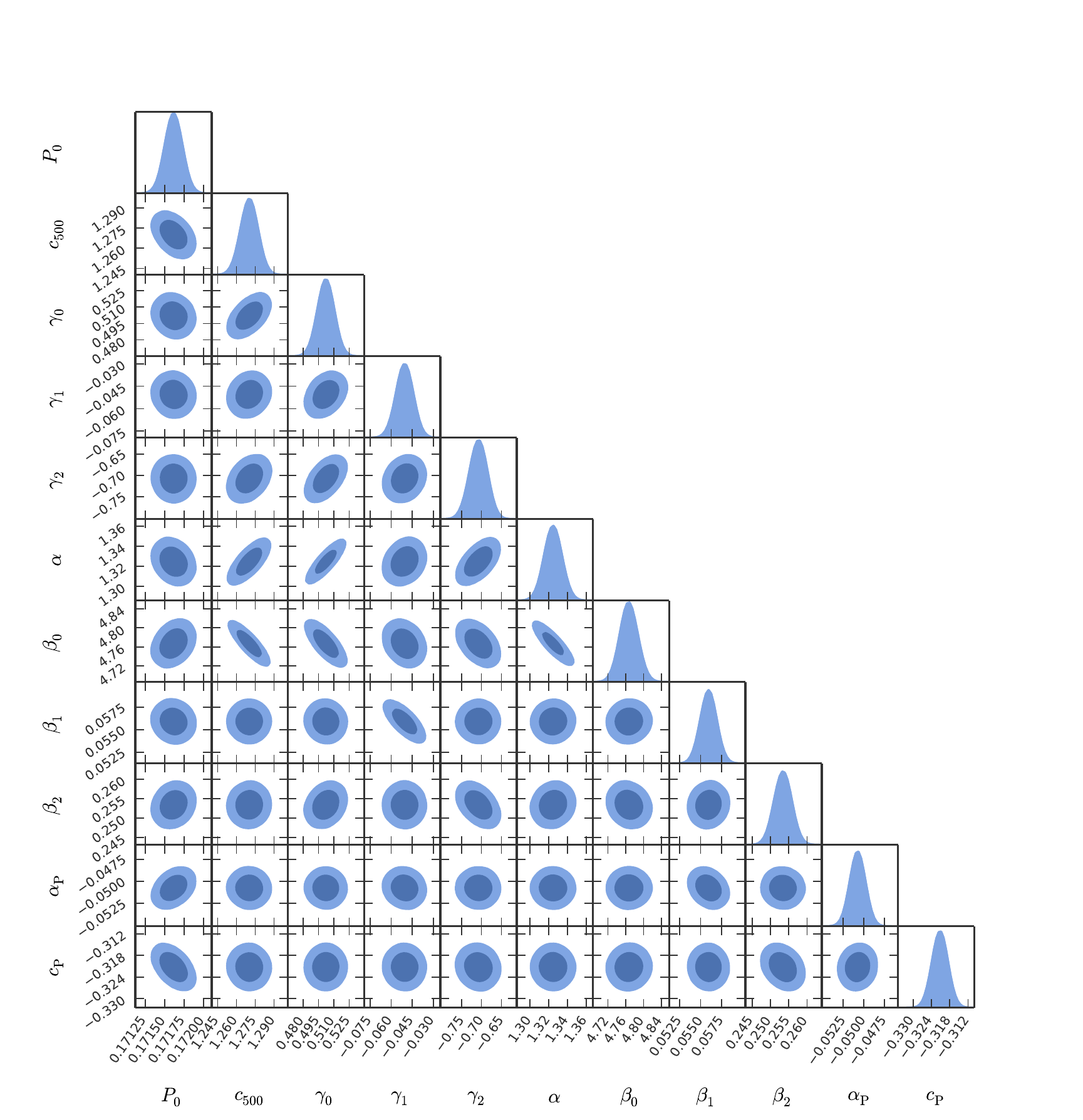}\hfil}
\vskip-0.15in
\caption[Constraints on the pressure profile]{Constraints on the e-GNFW model parameters from the simulated pressure profiles in the MPS. Shading from center to outside indicate the 1 and 2~$\sigma$ joint parameter confidence intervals, and the fully marginalized constraints for each parameter are at the right end of each row \citep{bocquet16triangle}.}
\label{fig:ysz-pressure-lik}
\end{figure*}
\begin{figure*}
\includegraphics[width= \textwidth]{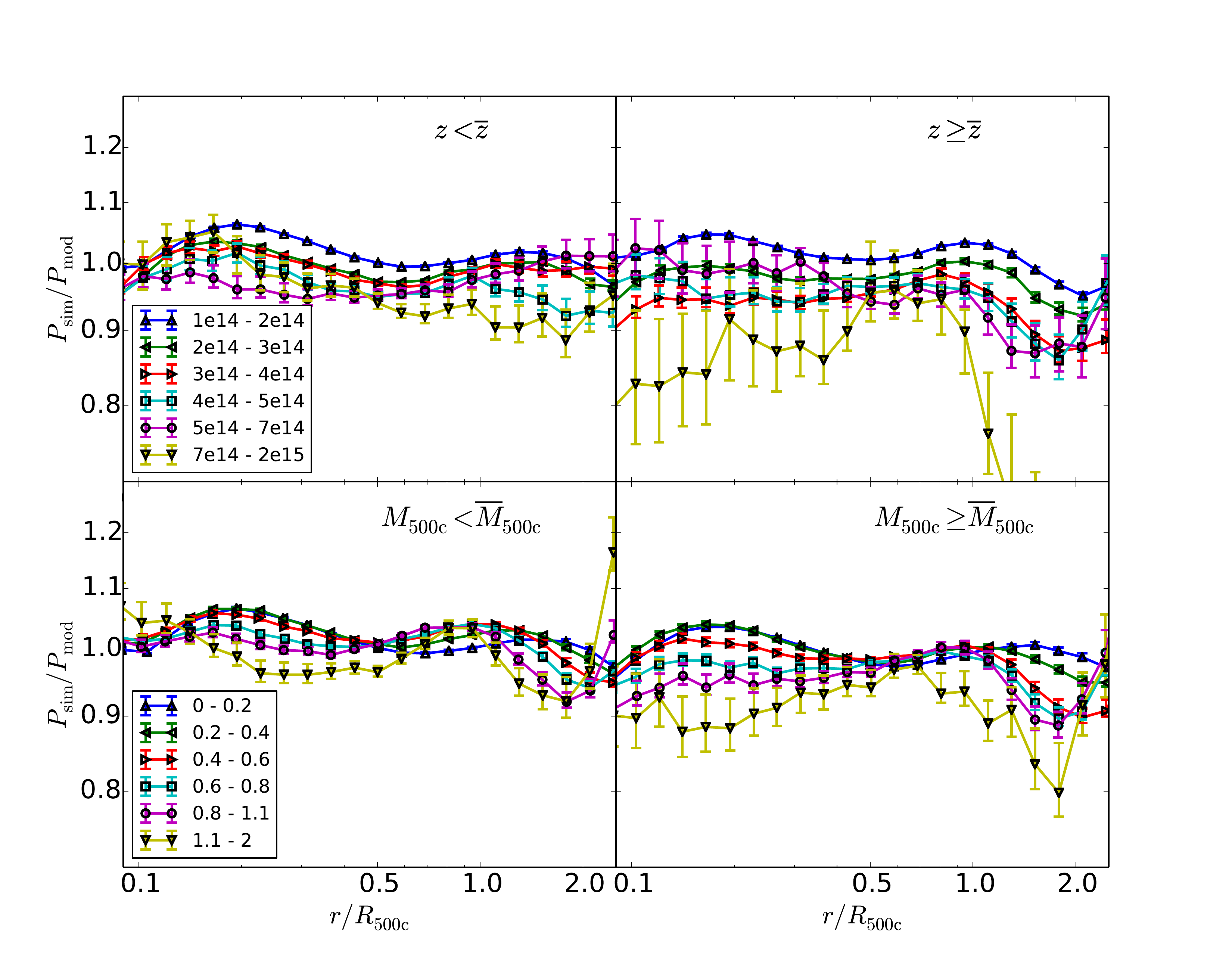}
\vskip-0.1in
\caption[The pressure profile trends]{Cluster pressure comparison in bins of mass ($M_{500\rm c}$, upper panels) and redshift (lower panels) with respect to the best fit e-GNFW profile (see Fig.~\ref{fig:pressure-trends} for further details). Clearly, the extended model is a much better fit to the whole range of clusters in the sample.}
\label{fig:extended-pressure-trends}
\end{figure*}
\subsection{Profile fitting}
In section~\ref{sec:press-mass-z-dependence}, we find clear trends in the shape of the pressure profile with cluster mass and redshift. The variation is most significant in the inner and outer regions of clusters. This motivates the modification of the model by including the mass and redshift dependencies to the inner and outer slopes of the GNFW profile to get an extended or e-GNFW profile written as
\BE \label{eq:ext-ysz-gnfw}
  P_\text{mod}(r,M,z) = P_{500}(M,z)\frac{c_{500}^{\gamma^{\prime}}(1+c_{500}^\alpha)^{(\beta^\prime-\gamma^\prime)/\alpha}}{(c_{500}~x)^{\gamma^{\prime}}[1+(c_{500}~x)^\alpha]^{(\beta^\prime-\gamma^\prime)/\alpha}},
\EE 
where $\gamma^\prime$ and $\beta^\prime$ are the modified inner and outer slopes
\BEA \label{ext-gamma-beta}
 \gamma^\prime = \gamma_0 \Big[\frac{\mfive}{3\times10^{14}\msun}\Big]^{\gamma_1}E(z)^{\gamma_2}, \nonumber \\
 \beta^\prime = \beta_0 \Big[\frac{\mfive}{3\times10^{14}\msun}\Big]^{\beta_1}E(z)^{\beta_2}.
\EEA

Thus, the e-GNFW profile is parametrized with 4 more parameters as compared to the GNFW model. 
We fit this extended model to the pressure profile from our simulations, taking advantage of the statistical power of MPS to constrain the complex high dimensional parameter space using the same methodology as described in section~\ref{sec:press-prof-fit}. We verify that the intrinsic scatter deduced from the most likely extended model is consistent with the intrinsic scatter shown in Fig.~\ref{fig:ysz-pressure-individual}. The best fit parameters along with 1-$\sigma$ uncertainties are presented in Table~\ref{tab:extended-ysz-press}. 

The marginalized posteriors are presented in Fig.~\ref{fig:ysz-pressure-lik} in the form of a triangle plot, which shows the joint confidence contours for different parameter pairs as well as the fully marginalized constraints for each single parameter. Interestingly, the additional covariances do not degrade the constraints on our model as the size of the cluster sample is very large.

\citet{battaglia12a} use a similar functional form to fit pressure profiles for a set of simulated clusters and show similar trends in $\beta$ with mass ($\beta_1=0.039$ for $z=0$ clusters) but find larger evolution with redshift ($\beta_2=0.415$ for $1.1\times 10^{14} \rm M_{\odot}<M_{\rm 200 c}<1.7\times 10^{14} \rm M_{\odot}$) as compared to our findings.
In an observational work, \citet{sayers16} measure \ac{sze} signal toward a set of 47 clusters with a median mass of $9.5\times 10^{14} \rm M_{\odot}$ and a median redshift of 0.4 using data from Planck and the ground-based Bolocam receiver. They find $\beta_1=0.077\pm 0.026$, which is consistent with our results. However, they find no evolution in $\beta$ with redshift, which as they imply, could be a result of sample selection, as their sample is biased toward relaxed cool-core systems at low-$z$ and toward disturbed merging systems at high-$z$.  We caution the reader while interpreting these results as different set of parameters are varied in these studies.

%
\begin{figure*}
\includegraphics[width= \textwidth]{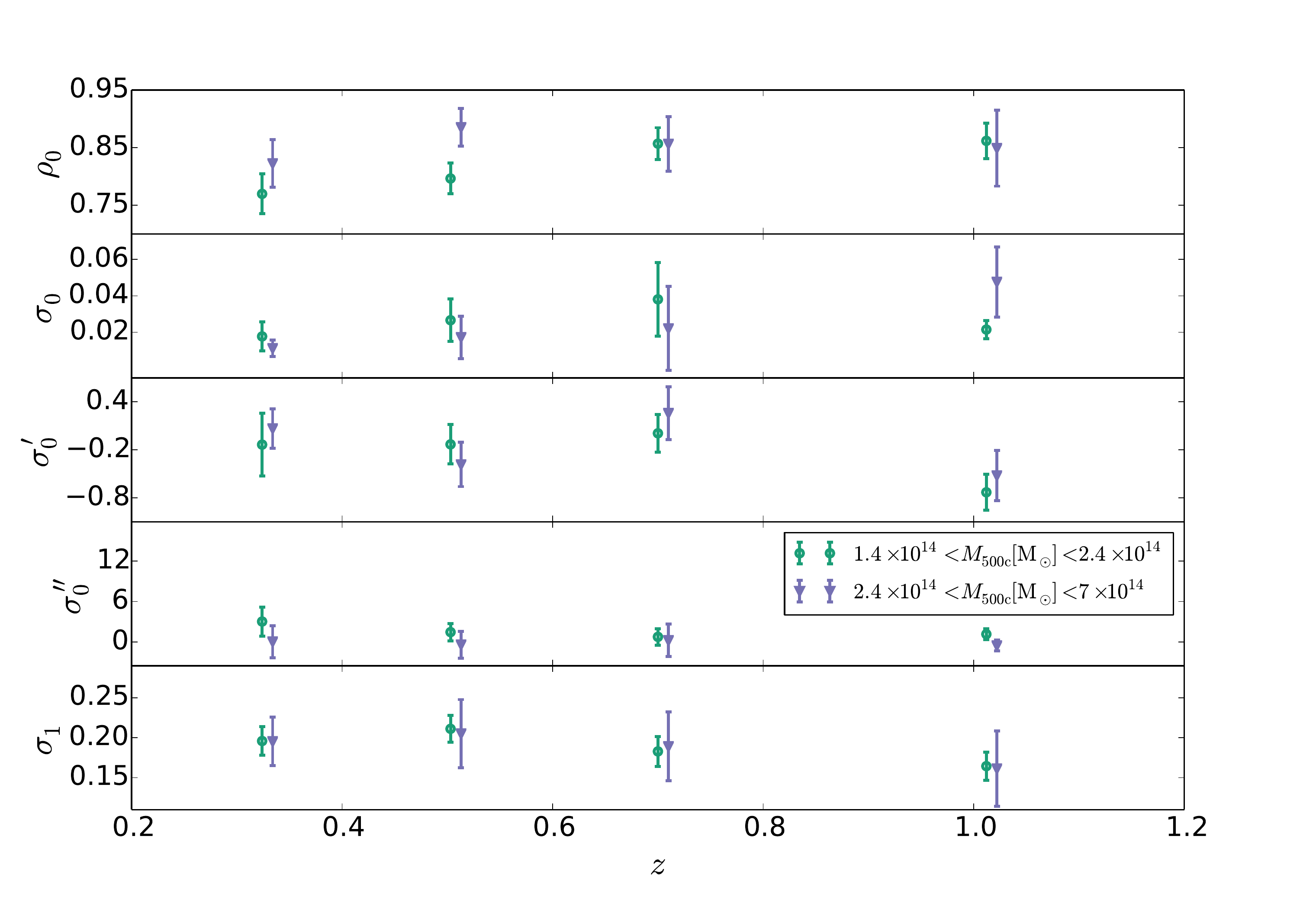}
\vskip-0.15in
\caption{Offset trends showing the consistency of the parameters with mass and redshift.}
\label{fig:extended-offset-trends}
\end{figure*}
\subsection{Variations in extended profile shape with mass and redshift}
Next we re-examine the systematic shape variations in pressure profiles with cluster mass and redshift and with respect to the e-GNFW model presented here. The clusters are selected from the full simulation box in the same manner as described in section~\ref{sec:press-mass-z-dependence}. The ratios between the pressure and the best fit e-GNFW model are shown in Fig.~\ref{fig:extended-pressure-trends} for different sub-samples. For cluster sub-sample with $z < \overline{z}$ (upper left panel), the model is within 5 percent of the simulations for the whole radial range and different mass bins. The sample with $z < \overline{z}$ is also consistent with best fit model within 10 percent (upper right panel), except for the last mass bin ($7\times 10^{14}<M_{500 \rm c}< 2\times 10^{14}$), which is because of very small number of clusters (25 clusters) in this range that has a small weight in the best fit model. Similarily, the model is consistent with clusters in different redshift bins, except for the last redshift bin ($1.1<z<2$) in the lower right panel. 

\begin{table}
\centering
  \caption[Constraints on e-GNFW Pressure Profile]{Constraints on the e-GNFW model parameters from fits to the simulated pressure profiles. These results are from clusters in the full simulation box with <$\mfive$>$=2.3\times10^{14}\msun$ and <$z$>$=0.31$. }
  \begin{tabular}{lr}
\hline
Parameter & values \\
\hline
$P_0$ & $0.1716\pm 0.0001$ \\
$c_{500}$ & $1.270\pm 0.006$ \\
$\gamma_0$ & $0.502\pm 0.008$ \\
$\gamma_1$ & $-0.050\pm 0.005$ \\
$\gamma_2$ & $-0.71\pm 0.02$ \\
$\alpha$ & $1.33\pm 0.01$ \\
$\beta_0$ & $4.77\pm 0.02$ \\
$\beta_1$ & $0.056\pm 0.001$ \\
$\beta_2$ & $0.254\pm 0.002$ \\
$\alpha_\text{P}$ & $-0.051\pm 0.001$ \\
$c_\text{P}$  & $-0.321\pm 0.002$ \\
\hline
  \end{tabular}
\label{tab:extended-ysz-press}
\end{table}
\begin{table}
\centering
\caption{The best fit parameters of the extended double Rayleigh function (see equation~(\ref{eq:rev-center-off})) fit to the radial offset distribution between the gravitational potential center and the $\ysz$ peak for clusters in the in the light cones for a revised model.}
\begin{tabular}{lr}
\hline
Parameter & \multicolumn{1}{c}{Light cone} \\ 
\hline
$\rho_0$  									&	$0.769\pm 0.035$	\\
$\sigma_0$									&	$0.018\pm 0.008$\\
$\sigma_0^\prime$					&	$-0.14\pm 0.39$\\
$\sigma_0^{\prime \prime}$		&	$3.037\pm 2.137$\\
$\sigma_1$									&	$0.195\pm 0.017$\\
\hline
\end{tabular}
\label{tab:ext-central-offset}
\end{table}
\section{Additional offset modeling}
\label{sec:revised-offset-model}
In section~\ref{sec:centroid-offset}, we described the variation in the offset distributions with mass and redshift and concluded that the relaxed cluster population show trends in the width of the Rayleigh distribution both with mass and redshift. In order to contain this evolution within the model, we extend the Rayleigh function by including mass and redshift dependencies to the width of the first component of the distribution as
\BE
P(x) = 2\pi x \left( \frac{\rho_0}{2 \pi \sigma_{\rm R}^2}e^{{-\frac{x^2}{2 \sigma_{\rm R}^2}} } + \frac{1- \rho_0}{2 \pi \sigma_1^2}e^{{-\frac{x^2}{2 \sigma_1^2}} } \right),
\label{eq:rev-center-off}
\EE
where $\sigma_{\rm R}$ is the extended width written as
\BE
\sigma_{\rm R}=\sigma_0 \Big[\frac{\mfive}{3\times10^{14}\msun}\Big]^{\sigma_0^\prime}E(z)^{\sigma_0^{\prime \prime}}.
\label{eq:sigma0-ext}
\EE
The best fit model parameters and 1-$\sigma$ uncertainties are presented in Table~\ref{tab:ext-central-offset} for all clusters in the light cone.

Next, we study the variation in the revised model parameters in the same range of cluster masses and redshifts as examined in section~\ref{sec:centroid-offset}. The best fit parameters and uncertainties are shown in Fig.~\ref{fig:extended-offset-trends}, which clearly indicates that the trends in the width of the relaxed population is captured within the framework of this extended model. 

\bibliographystyle{mnras}
\bibliography{Magneticum_SZE}

\begin{thebibliography}{}
\makeatletter
\relax
\def\mn@urlcharsother{\let\do\@makeother \do\$\do\&\do\#\do\^\do\_\do\%\do\~}
\def\mn@doi{\begingroup\mn@urlcharsother \@ifnextchar [ {\mn@doi@}
  {\mn@doi@[]}}
\def\mn@doi@[#1]#2{\def\@tempa{#1}\ifx\@tempa\@empty \href
  {http://dx.doi.org/#2} {doi:#2}\else \href {http://dx.doi.org/#2} {#1}\fi
  \endgroup}
\def\mn@eprint#1#2{\mn@eprint@#1:#2::\@nil}
\def\mn@eprint@arXiv#1{\href {http://arxiv.org/abs/#1} {{\tt arXiv:#1}}}
\def\mn@eprint@dblp#1{\href {http://dblp.uni-trier.de/rec/bibtex/#1.xml}
  {dblp:#1}}
\def\mn@eprint@#1:#2:#3:#4\@nil{\def\@tempa {#1}\def\@tempb {#2}\def\@tempc
  {#3}\ifx \@tempc \@empty \let \@tempc \@tempb \let \@tempb \@tempa \fi \ifx
  \@tempb \@empty \def\@tempb {arXiv}\fi \@ifundefined
  {mn@eprint@\@tempb}{\@tempb:\@tempc}{\expandafter \expandafter \csname
  mn@eprint@\@tempb\endcsname \expandafter{\@tempc}}}

\bibitem[\protect\citeauthoryear{{Arnaud}, {Pratt}, {Piffaretti},
  {B{\"o}hringer}, {Croston}  \& {Pointecouteau}}{{Arnaud}
  et~al.}{2010}]{arnaud10}
{Arnaud} M.,  {Pratt} G.~W.,  {Piffaretti} R.,  {B{\"o}hringer} H.,  {Croston}
  J.~H.,   {Pointecouteau} E.,  2010, \mn@doi [\aap]
  {10.1051/0004-6361/200913416}, \href
  {http://adsabs.harvard.edu/abs/2010A%26A...517A..92A} {517, A92}

\bibitem[\protect\citeauthoryear{{Arthur} et~al.,}{{Arthur}
  et~al.}{2017}]{arthur16}
{Arthur} J.,  et~al., 2017, \mn@doi [\mnras] {10.1093/mnras/stw2424}, \href
  {http://adsabs.harvard.edu/abs/2017MNRAS.464.2027A} {464, 2027}

\bibitem[\protect\citeauthoryear{{Avestruz}, {Nagai}  \& {Lau}}{{Avestruz}
  et~al.}{2016}]{avestruz16}
{Avestruz} C.,  {Nagai} D.,   {Lau} E.~T.,  2016, preprint, \href
  {http://adsabs.harvard.edu/abs/2016arXiv160501723A} {} (\mn@eprint {arXiv}
  {1605.01723})

\bibitem[\protect\citeauthoryear{{Battaglia}, {Bond}, {Pfrommer}, {Sievers}  \&
  {Sijacki}}{{Battaglia} et~al.}{2010}]{battaglia10}
{Battaglia} N.,  {Bond} J.~R.,  {Pfrommer} C.,  {Sievers} J.~L.,   {Sijacki}
  D.,  2010, \mn@doi [\apj] {10.1088/0004-637X/725/1/91}, \href
  {http://adsabs.harvard.edu/abs/2010ApJ...725...91B} {725, 91}

\bibitem[\protect\citeauthoryear{{Battaglia}, {Bond}, {Pfrommer}  \&
  {Sievers}}{{Battaglia} et~al.}{2011}]{battaglia11}
{Battaglia} N.,  {Bond} J.~R.,  {Pfrommer} C.,   {Sievers} J.~L.,  2011,
  ArXiv:1109.3711, \href {http://adsabs.harvard.edu/abs/2011arXiv1109.3711B} {}

\bibitem[\protect\citeauthoryear{{Battaglia}, {Bond}, {Pfrommer}  \&
  {Sievers}}{{Battaglia} et~al.}{2012a}]{battaglia12}
{Battaglia} N.,  {Bond} J.~R.,  {Pfrommer} C.,   {Sievers} J.~L.,  2012a,
  \mn@doi [\apj] {10.1088/0004-637X/758/2/74}, \href
  {http://adsabs.harvard.edu/abs/2012ApJ...758...74B} {758, 74}

\bibitem[\protect\citeauthoryear{{Battaglia}, {Bond}, {Pfrommer}  \&
  {Sievers}}{{Battaglia} et~al.}{2012b}]{battaglia12a}
{Battaglia} N.,  {Bond} J.~R.,  {Pfrommer} C.,   {Sievers} J.~L.,  2012b,
  \mn@doi [\apj] {10.1088/0004-637X/758/2/75}, \href
  {http://adsabs.harvard.edu/abs/2012ApJ...758...75B} {758, 75}

\bibitem[\protect\citeauthoryear{{Beck} et~al.,}{{Beck} et~al.}{2016}]{beck16}
{Beck} A.~M.,  et~al., 2016, \mn@doi [\mnras] {10.1093/mnras/stv2443}, \href
  {http://adsabs.harvard.edu/abs/2016MNRAS.455.2110B} {455, 2110}

\bibitem[\protect\citeauthoryear{{Biesiadzinski}, {McMahon}, {Miller}, {Nord}
  \& {Shaw}}{{Biesiadzinski} et~al.}{2012}]{Biesiadzinski12}
{Biesiadzinski} T.,  {McMahon} J.,  {Miller} C.~J.,  {Nord} B.,   {Shaw} L.,
  2012, \mn@doi [\apj] {10.1088/0004-637X/757/1/1}, \href
  {http://adsabs.harvard.edu/abs/2012ApJ...757....1B} {757, 1}

\bibitem[\protect\citeauthoryear{{Biffi} et~al.,}{{Biffi}
  et~al.}{2016}]{biffi16}
{Biffi} V.,  et~al., 2016, \mn@doi [\apj] {10.3847/0004-637X/827/2/112}, \href
  {http://adsabs.harvard.edu/abs/2016ApJ...827..112B} {827, 112}

\bibitem[\protect\citeauthoryear{{Birkinshaw}, {Gull}  \&
  {Hardebeck}}{{Birkinshaw} et~al.}{1984}]{birkinshaw84}
{Birkinshaw} M.,  {Gull} S.~F.,   {Hardebeck} H.,  1984, \mn@doi [\nat]
  {10.1038/309034a0}, \href {http://adsabs.harvard.edu/abs/1984Natur.309...34B}
  {309, 34}

\bibitem[\protect\citeauthoryear{{Bleem} et~al.,}{{Bleem}
  et~al.}{2015}]{bleem15}
{Bleem} L.~E.,  et~al., 2015, \mn@doi [\apjs] {10.1088/0067-0049/216/2/27},
  \href {http://adsabs.harvard.edu/abs/2015ApJS..216...27B} {216, 27}

\bibitem[\protect\citeauthoryear{Bocquet \& Carter}{Bocquet \&
  Carter}{2016}]{bocquet16triangle}
Bocquet S.,  Carter F.~W.,  2016, \mn@doi [The Journal of Open Source Software]
  {10.21105/joss.00046}, 1

\bibitem[\protect\citeauthoryear{{Bocquet} et~al.,}{{Bocquet}
  et~al.}{2015}]{bocquet15}
{Bocquet} S.,  et~al., 2015, \mn@doi [\apj] {10.1088/0004-637X/799/2/214},
  \href {http://adsabs.harvard.edu/abs/2015ApJ...799..214B} {799, 214}

\bibitem[\protect\citeauthoryear{{Bocquet}, {Saro}, {Dolag}  \&
  {Mohr}}{{Bocquet} et~al.}{2016}]{bocquet15b}
{Bocquet} S.,  {Saro} A.,  {Dolag} K.,   {Mohr} J.~J.,  2016, \mn@doi [\mnras]
  {10.1093/mnras/stv2657}, \href
  {http://adsabs.harvard.edu/abs/2016MNRAS.456.2361B} {456, 2361}

\bibitem[\protect\citeauthoryear{{B{\"o}hringer} et~al.,}{{B{\"o}hringer}
  et~al.}{2007}]{boehringer07}
{B{\"o}hringer} H.,  et~al., 2007, \mn@doi [\aap] {10.1051/0004-6361:20066740},
  \href {http://adsabs.harvard.edu/abs/2007A%26A...469..363B} {469, 363}

\bibitem[\protect\citeauthoryear{{Bonaldi}, {Tormen}, {Dolag}  \&
  {Moscardini}}{{Bonaldi} et~al.}{2007}]{bonaldi07}
{Bonaldi} A.,  {Tormen} G.,  {Dolag} K.,   {Moscardini} L.,  2007, \mn@doi
  [\mnras] {10.1111/j.1365-2966.2007.11893.x}, \href
  {http://adsabs.harvard.edu/abs/2007MNRAS.378.1248B} {378, 1248}

\bibitem[\protect\citeauthoryear{{Borgani} et~al.,}{{Borgani}
  et~al.}{2004}]{borgani04}
{Borgani} S.,  et~al., 2004, \mn@doi [\mnras]
  {10.1111/j.1365-2966.2004.07431.x}, \href
  {http://adsabs.harvard.edu/abs/2004MNRAS.348.1078B} {348, 1078}

\bibitem[\protect\citeauthoryear{{Burns}, {Skillman}  \& {O'Shea}}{{Burns}
  et~al.}{2010}]{burn10}
{Burns} J.~O.,  {Skillman} S.~W.,   {O'Shea} B.~W.,  2010, \mn@doi [\apj]
  {10.1088/0004-637X/721/2/1105}, \href
  {http://adsabs.harvard.edu/abs/2010ApJ...721.1105B} {721, 1105}

\bibitem[\protect\citeauthoryear{{Carlstrom}, {Holder}  \& {Reese}}{{Carlstrom}
  et~al.}{2002}]{carlstrom02}
{Carlstrom} J.~E.,  {Holder} G.~P.,   {Reese} E.~D.,  2002, \araa, \href
  {http://adsabs.harvard.edu/cgi-bin/nph-bib_query?bibcode=2002ARA%26A..40..643C&amp;db_key=AST}
  {40, 643}

\bibitem[\protect\citeauthoryear{{Carlstrom} et~al.,}{{Carlstrom}
  et~al.}{2011}]{carlstrom11}
{Carlstrom} J.~E.,  et~al., 2011, \mn@doi [\pasp] {10.1086/659879}, \href
  {http://adsabs.harvard.edu/abs/2011PASP..123..568C} {123, 568}

\bibitem[\protect\citeauthoryear{{Chiu} \& {Molnar}}{{Chiu} \&
  {Molnar}}{2012}]{chiu12}
{Chiu} I.-N.~T.,  {Molnar} S.~M.,  2012, \mn@doi [\apj]
  {10.1088/0004-637X/756/1/1}, \href
  {http://adsabs.harvard.edu/abs/2012ApJ...756....1C} {756, 1}

\bibitem[\protect\citeauthoryear{{Cui} et~al.,}{{Cui} et~al.}{2016}]{cui16}
{Cui} W.,  et~al., 2016, \mn@doi [\mnras] {10.1093/mnras/stw603}, \href
  {http://adsabs.harvard.edu/abs/2016MNRAS.458.4052C} {458, 4052}

\bibitem[\protect\citeauthoryear{{DES Collaboration}}{{DES
  Collaboration}}{2005}]{DES05}
{DES Collaboration} 2005, ArXiv Astrophysics e-prints, \href
  {http://adsabs.harvard.edu/abs/2005astro.ph.10346T} {}

\bibitem[\protect\citeauthoryear{{Da Silva}, {Kay}, {Liddle}, {Thomas},
  {Pearce}  \& {Barbosa}}{{Da Silva} et~al.}{2001}]{desilva00}
{Da Silva} A.~C.,  {Kay} S.~T.,  {Liddle} A.~R.,  {Thomas} P.~A.,  {Pearce}
  F.~R.,   {Barbosa} D.,  2001, \mn@doi [\apjl] {10.1086/324574}, \href
  {http://adsabs.harvard.edu/abs/2001ApJ...561L..15D} {561, L15}

\bibitem[\protect\citeauthoryear{{Davis}, {Efstathiou}, {Frenk}  \&
  {White}}{{Davis} et~al.}{1985}]{davis85}
{Davis} M.,  {Efstathiou} G.,  {Frenk} C.~S.,   {White} S.~D.~M.,  1985,
  \mn@doi [\apj] {10.1086/163168}, \href
  {http://adsabs.harvard.edu/abs/1985ApJ...292..371D} {292, 371}

\bibitem[\protect\citeauthoryear{{Dehnen} \& {Aly}}{{Dehnen} \&
  {Aly}}{2012}]{dehnen12}
{Dehnen} W.,  {Aly} H.,  2012, \mn@doi [\mnras]
  {10.1111/j.1365-2966.2012.21439.x}, \href
  {http://adsabs.harvard.edu/abs/2012MNRAS.425.1068D} {425, 1068}

\bibitem[\protect\citeauthoryear{{Dolag} \& {Stasyszyn}}{{Dolag} \&
  {Stasyszyn}}{2009}]{dolag09}
{Dolag} K.,  {Stasyszyn} F.,  2009, \mn@doi [\mnras]
  {10.1111/j.1365-2966.2009.15181.x}, \href
  {http://adsabs.harvard.edu/abs/2009MNRAS.398.1678D} {398, 1678}

\bibitem[\protect\citeauthoryear{{Dolag}, {Jubelgas}, {Springel}, {Borgani}  \&
  {Rasia}}{{Dolag} et~al.}{2004}]{dolag04a}
{Dolag} K.,  {Jubelgas} M.,  {Springel} V.,  {Borgani} S.,   {Rasia} E.,  2004,
  \mn@doi [\apjl] {10.1086/420966}, \href
  {http://adsabs.harvard.edu/abs/2004ApJ...606L..97D} {606, L97}

\bibitem[\protect\citeauthoryear{{Dolag}, {Hansen}, {Roncarelli}  \&
  {Moscardini}}{{Dolag} et~al.}{2005a}]{dolag05}
{Dolag} K.,  {Hansen} F.~K.,  {Roncarelli} M.,   {Moscardini} L.,  2005a,
  \mn@doi [\mnras] {10.1111/j.1365-2966.2005.09452.x}, \href
  {http://adsabs.harvard.edu/abs/2005MNRAS.363...29D} {363, 29}

\bibitem[\protect\citeauthoryear{{Dolag}, {Vazza}, {Brunetti}  \&
  {Tormen}}{{Dolag} et~al.}{2005b}]{dolag05s}
{Dolag} K.,  {Vazza} F.,  {Brunetti} G.,   {Tormen} G.,  2005b, \mn@doi
  [\mnras] {10.1111/j.1365-2966.2005.09630.x}, \href
  {http://adsabs.harvard.edu/abs/2005MNRAS.364..753D} {364, 753}

\bibitem[\protect\citeauthoryear{{Dolag}, {Borgani}, {Murante}  \&
  {Springel}}{{Dolag} et~al.}{2009}]{dolag09a}
{Dolag} K.,  {Borgani} S.,  {Murante} G.,   {Springel} V.,  2009, \mn@doi
  [\mnras] {10.1111/j.1365-2966.2009.15034.x}, \href
  {http://adsabs.harvard.edu/abs/2009MNRAS.399..497D} {399, 497}

\bibitem[\protect\citeauthoryear{{Dolag}, {Komatsu}  \& {Sunyaev}}{{Dolag}
  et~al.}{2016a}]{dolag15}
{Dolag} K.,  {Komatsu} E.,   {Sunyaev} R.,  2016a, \mn@doi [\mnras]
  {10.1093/mnras/stw2035}, \href
  {http://adsabs.harvard.edu/abs/2016MNRAS.463.1797D} {463, 1797}

\bibitem[\protect\citeauthoryear{{Dolag}, {Komatsu}  \& {Sunyaev}}{{Dolag}
  et~al.}{2016b}]{dolag16}
{Dolag} K.,  {Komatsu} E.,   {Sunyaev} R.,  2016b, \mn@doi [\mnras]
  {10.1093/mnras/stw2035}, \href
  {http://adsabs.harvard.edu/abs/2016MNRAS.463.1797D} {463, 1797}

\bibitem[\protect\citeauthoryear{{Dolag et al. in}}{{Dolag et al.
  in}}{prep}]{dolag13}
{Dolag et al. in} prep., preprint (\mn@eprint {arXiv} {9999.9999})

\bibitem[\protect\citeauthoryear{{Donnert}, {Dolag}, {Brunetti}  \&
  {Cassano}}{{Donnert} et~al.}{2013}]{donnert13}
{Donnert} J.,  {Dolag} K.,  {Brunetti} G.,   {Cassano} R.,  2013, \mn@doi
  [\mnras] {10.1093/mnras/sts628}, \href
  {http://adsabs.harvard.edu/abs/2013MNRAS.429.3564D} {429, 3564}

\bibitem[\protect\citeauthoryear{{Du} \& {Fan}}{{Du} \& {Fan}}{2014}]{du14}
{Du} W.,  {Fan} Z.,  2014, \mn@doi [\apj] {10.1088/0004-637X/785/1/57}, \href
  {http://adsabs.harvard.edu/abs/2014ApJ...785...57D} {785, 57}

\bibitem[\protect\citeauthoryear{{Eisenstein} \& {Hu}}{{Eisenstein} \&
  {Hu}}{1998}]{eisenstein98}
{Eisenstein} D.~J.,  {Hu} W.,  1998, \mn@doi [\apj] {10.1086/305424}, \href
  {http://adsabs.harvard.edu/abs/1998ApJ...496..605E} {496, 605}

\bibitem[\protect\citeauthoryear{{Elahi} et~al.,}{{Elahi}
  et~al.}{2016}]{elahi16}
{Elahi} P.~J.,  et~al., 2016, \mn@doi [\mnras] {10.1093/mnras/stw338}, \href
  {http://adsabs.harvard.edu/abs/2016MNRAS.458.1096E} {458, 1096}

\bibitem[\protect\citeauthoryear{{Ensslin}, {Biermann}, {Kronberg}  \&
  {Wu}}{{Ensslin} et~al.}{1997}]{ensslin97}
{Ensslin} T.~A.,  {Biermann} P.~L.,  {Kronberg} P.~P.,   {Wu} X.-P.,  1997,
  \apj, \href {http://adsabs.harvard.edu/abs/1997ApJ...477..560E} {477, 560}

\bibitem[\protect\citeauthoryear{{Evrard}}{{Evrard}}{1990}]{evrard90}
{Evrard} A.~E.,  1990, \apj, 363, 349

\bibitem[\protect\citeauthoryear{{Fabjan}, {Tornatore}, {Borgani}, {Saro}  \&
  {Dolag}}{{Fabjan} et~al.}{2008}]{fabjan08}
{Fabjan} D.,  {Tornatore} L.,  {Borgani} S.,  {Saro} A.,   {Dolag} K.,  2008,
  \mn@doi [\mnras] {10.1111/j.1365-2966.2008.13122.x}, \href
  {http://adsabs.harvard.edu/abs/2008MNRAS.386.1265F} {386, 1265}

\bibitem[\protect\citeauthoryear{{Fabjan}, {Borgani}, {Tornatore}, {Saro},
  {Murante}  \& {Dolag}}{{Fabjan} et~al.}{2010}]{fabjan10}
{Fabjan} D.,  {Borgani} S.,  {Tornatore} L.,  {Saro} A.,  {Murante} G.,
  {Dolag} K.,  2010, \mn@doi [\mnras] {10.1111/j.1365-2966.2009.15794.x}, \href
  {http://adsabs.harvard.edu/abs/2010MNRAS.401.1670F} {401, 1670}

\bibitem[\protect\citeauthoryear{{Fang}, {Humphrey}  \& {Buote}}{{Fang}
  et~al.}{2009}]{fang09}
{Fang} T.,  {Humphrey} P.,   {Buote} D.,  2009, \mn@doi [\apj]
  {10.1088/0004-637X/691/2/1648}, \href
  {http://adsabs.harvard.edu/abs/2009ApJ...691.1648F} {691, 1648}

\bibitem[\protect\citeauthoryear{{Foreman-Mackey}, {Hogg}, {Lang}  \&
  {Goodman}}{{Foreman-Mackey} et~al.}{2013}]{mackey13}
{Foreman-Mackey} D.,  {Hogg} D.~W.,  {Lang} D.,   {Goodman} J.,  2013, \mn@doi
  [\pasp] {10.1086/670067}, \href
  {http://adsabs.harvard.edu/abs/2013PASP..125..306F} {125, 306}

\bibitem[\protect\citeauthoryear{{Gazzola} \& {Pearce}}{{Gazzola} \&
  {Pearce}}{2007}]{gazzola07}
{Gazzola} L.,  {Pearce} F.~R.,  2007, in {B{\"o}hringer} H.,  {Pratt} G.~W.,
  {Finoguenov} A.,   {Schuecker} P.,  eds, Heating versus Cooling in Galaxies
  and Clusters of Galaxies. p.~412 (\mn@eprint {} {astro-ph/0611715}),
  \mn@doi{10.1007/978-3-540-73484-0_76}

\bibitem[\protect\citeauthoryear{{George} et~al.,}{{George}
  et~al.}{2012}]{george12a}
{George} M.~R.,  et~al., 2012, \mn@doi [\apj] {10.1088/0004-637X/757/1/2},
  \href {http://adsabs.harvard.edu/abs/2012ApJ...757....2G} {757, 2}

\bibitem[\protect\citeauthoryear{{Gottloeber}, {Yepes}, {Wagner}  \&
  {Sevilla}}{{Gottloeber} et~al.}{2006}]{gottloaeber06}
{Gottloeber} S.,  {Yepes} G.,  {Wagner} C.,   {Sevilla} R.,  2006, ArXiv
  Astrophysics e-prints, \href
  {http://adsabs.harvard.edu/abs/2006astro.ph..8289G} {}

\bibitem[\protect\citeauthoryear{{Gupta} et~al.,}{{Gupta}
  et~al.}{2016}]{gupta16}
{Gupta} N.,  et~al., 2016, preprint, \href
  {http://adsabs.harvard.edu/abs/2016arXiv160505329G} {} (\mn@eprint {arXiv}
  {1605.05329})

\bibitem[\protect\citeauthoryear{{Haiman}, {Mohr}  \& {Holder}}{{Haiman}
  et~al.}{2001}]{haiman01}
{Haiman} Z.,  {Mohr} J.~J.,   {Holder} G.~P.,  2001, \apj, 553, 545

\bibitem[\protect\citeauthoryear{{Hasselfield} et~al.,}{{Hasselfield}
  et~al.}{2013}]{hasselfield13}
{Hasselfield} M.,  et~al., 2013, \mn@doi [\jcap]
  {10.1088/1475-7516/2013/07/008}, \href
  {http://adsabs.harvard.edu/abs/2013JCAP...07..008H} {7, 8}

\bibitem[\protect\citeauthoryear{{Hirschmann}, {Dolag}, {Saro}, {Bachmann},
  {Borgani}  \& {Burkert}}{{Hirschmann} et~al.}{2014}]{hirschmann14}
{Hirschmann} M.,  {Dolag} K.,  {Saro} A.,  {Bachmann} L.,  {Borgani} S.,
  {Burkert} A.,  2014, \mn@doi [\mnras] {10.1093/mnras/stu1023}, \href
  {http://adsabs.harvard.edu/abs/2014MNRAS.442.2304H} {442, 2304}

\bibitem[\protect\citeauthoryear{{Hoekstra}, {Herbonnet}, {Muzzin}, {Babul},
  {Mahdavi}, {Viola}  \& {Cacciato}}{{Hoekstra} et~al.}{2015}]{hoekstra15}
{Hoekstra} H.,  {Herbonnet} R.,  {Muzzin} A.,  {Babul} A.,  {Mahdavi} A.,
  {Viola} M.,   {Cacciato} M.,  2015, \mn@doi [\mnras] {10.1093/mnras/stv275},
  \href {http://adsabs.harvard.edu/abs/2015MNRAS.449..685H} {449, 685}

\bibitem[\protect\citeauthoryear{{Johnston} et~al.,}{{Johnston}
  et~al.}{2007}]{johnston07}
{Johnston} D.~E.,  et~al., 2007, preprint, \href
  {http://adsabs.harvard.edu/abs/2007arXiv0709.1159J} {} (\mn@eprint {arXiv}
  {0709.1159})

\bibitem[\protect\citeauthoryear{{Kaiser} \& {Silk}}{{Kaiser} \&
  {Silk}}{1986}]{kaiser86}
{Kaiser} N.,  {Silk} J.,  1986, \nat, 324, 529

\bibitem[\protect\citeauthoryear{{Kay}, {Thomas}, {Jenkins}  \& {Pearce}}{{Kay}
  et~al.}{2004}]{kay04}
{Kay} S.~T.,  {Thomas} P.~A.,  {Jenkins} A.,   {Pearce} F.~R.,  2004, \mn@doi
  [\mnras] {10.1111/j.1365-2966.2004.08383.x}, \href
  {http://adsabs.harvard.edu/abs/2004MNRAS.355.1091K} {355, 1091}

\bibitem[\protect\citeauthoryear{{Kay}, {Peel}, {Short}, {Thomas}, {Young},
  {Battye}, {Liddle}  \& {Pearce}}{{Kay} et~al.}{2012}]{kay12}
{Kay} S.~T.,  {Peel} M.~W.,  {Short} C.~J.,  {Thomas} P.~A.,  {Young} O.~E.,
  {Battye} R.~A.,  {Liddle} A.~R.,   {Pearce} F.~R.,  2012, \mn@doi [\mnras]
  {10.1111/j.1365-2966.2012.20623.x}, \href
  {http://adsabs.harvard.edu/abs/2012MNRAS.422.1999K} {422, 1999}

\bibitem[\protect\citeauthoryear{{Komatsu} et~al.,}{{Komatsu}
  et~al.}{2011}]{komatsu11}
{Komatsu} E.,  et~al., 2011, \mn@doi [\apjs] {10.1088/0067-0049/192/2/18},
  \href {http://adsabs.harvard.edu/abs/2011ApJS..192...18K} {192, 18}

\bibitem[\protect\citeauthoryear{{Kravtsov}, {Nagai}  \&
  {Vikhlinin}}{{Kravtsov} et~al.}{2005}]{kravtsov05}
{Kravtsov} A.~V.,  {Nagai} D.,   {Vikhlinin} A.~A.,  2005, \mn@doi [\apj]
  {10.1086/429796}, \href
  {http://esoads.eso.org/cgi-bin/nph-bib_query?bibcode=2005ApJ...625..588K&db_key=AST}
  {625, 588}

\bibitem[\protect\citeauthoryear{{Kravtsov}, {Vikhlinin}  \&
  {Nagai}}{{Kravtsov} et~al.}{2006}]{kravtsov06}
{Kravtsov} A.~V.,  {Vikhlinin} A.,   {Nagai} D.,  2006, \mn@doi [\apj]
  {10.1086/506319}, \href {http://adsabs.harvard.edu/abs/2006ApJ...650..128K}
  {650, 128}

\bibitem[\protect\citeauthoryear{{LSST Science Collaboration}}{{LSST Science
  Collaboration}}{2009}]{LSST09}
{LSST Science Collaboration} 2009, preprint, \href
  {http://adsabs.harvard.edu/abs/2009arXiv0912.0201L} {} (\mn@eprint {arXiv}
  {0912.0201})

\bibitem[\protect\citeauthoryear{{Lau}, {Kravtsov}  \& {Nagai}}{{Lau}
  et~al.}{2009}]{lau09}
{Lau} E.~T.,  {Kravtsov} A.~V.,   {Nagai} D.,  2009, \mn@doi [\apj]
  {10.1088/0004-637X/705/2/1129}, \href
  {http://adsabs.harvard.edu/abs/2009ApJ...705.1129L} {705, 1129}

\bibitem[\protect\citeauthoryear{{Lau}, {Nagai}, {Avestruz}, {Nelson}  \&
  {Vikhlinin}}{{Lau} et~al.}{2015}]{lau15}
{Lau} E.~T.,  {Nagai} D.,  {Avestruz} C.,  {Nelson} K.,   {Vikhlinin} A.,
  2015, \mn@doi [\apj] {10.1088/0004-637X/806/1/68}, \href
  {http://adsabs.harvard.edu/abs/2015ApJ...806...68L} {806, 68}

\bibitem[\protect\citeauthoryear{{Laureijs} et~al.,}{{Laureijs}
  et~al.}{2011}]{laureijs11}
{Laureijs} R.,  et~al., 2011, preprint, \href
  {http://adsabs.harvard.edu/abs/2011arXiv1110.3193L} {} (\mn@eprint {arXiv}
  {1110.3193})

\bibitem[\protect\citeauthoryear{{Le Brun}, {McCarthy}, {Schaye}  \&
  {Ponman}}{{Le Brun} et~al.}{2014}]{lebrun14}
{Le Brun} A.~M.~C.,  {McCarthy} I.~G.,  {Schaye} J.,   {Ponman} T.~J.,  2014,
  \mn@doi [\mnras] {10.1093/mnras/stu608}, \href
  {http://adsabs.harvard.edu/abs/2014MNRAS.441.1270L} {441, 1270}

\bibitem[\protect\citeauthoryear{{Le Brun}, {McCarthy}, {Schaye}  \&
  {Ponman}}{{Le Brun} et~al.}{2016}]{lebrun16}
{Le Brun} A.~M.~C.,  {McCarthy} I.~G.,  {Schaye} J.,   {Ponman} T.~J.,  2016,
  preprint, \href {http://adsabs.harvard.edu/abs/2016arXiv160604545L} {}
  (\mn@eprint {arXiv} {1606.04545})

\bibitem[\protect\citeauthoryear{{Lin} \& {Mohr}}{{Lin} \&
  {Mohr}}{2004}]{lin04b}
{Lin} Y.-T.,  {Mohr} J.~J.,  2004, \mn@doi [\apj] {10.1086/425412}, \href
  {http://adsabs.harvard.edu/abs/2004ApJ...617..879L} {617, 879}

\bibitem[\protect\citeauthoryear{{McCarthy}, {Babul}, {Holder}  \&
  {Balogh}}{{McCarthy} et~al.}{2003}]{mccarthy03}
{McCarthy} I.~G.,  {Babul} A.,  {Holder} G.~P.,   {Balogh} M.~L.,  2003,
  \mn@doi [\apj] {10.1086/375486}, \href
  {http://adsabs.harvard.edu/abs/2003ApJ...591..515M} {591, 515}

\bibitem[\protect\citeauthoryear{{McDonald} et~al.,}{{McDonald}
  et~al.}{2014}]{mcdonald14b}
{McDonald} M.,  et~al., 2014, \mn@doi [\apj] {10.1088/0004-637X/794/1/67},
  \href {http://adsabs.harvard.edu/abs/2014ApJ...794...67M} {794, 67}

\bibitem[\protect\citeauthoryear{{Merloni} et~al.,}{{Merloni}
  et~al.}{2012}]{merloni12}
{Merloni} A.,  et~al., 2012, preprint, \href
  {http://adsabs.harvard.edu/abs/2012arXiv1209.3114M} {} (\mn@eprint {arXiv}
  {1209.3114})

\bibitem[\protect\citeauthoryear{Mohr \& Evrard}{Mohr \& Evrard}{1997}]{mohr97}
Mohr J.,  Evrard A.,  1997, \apj, 491, 38

\bibitem[\protect\citeauthoryear{{Mohr}, {Mathiesen}  \& {Evrard}}{{Mohr}
  et~al.}{1999}]{mohr99}
{Mohr} J.~J.,  {Mathiesen} B.,   {Evrard} A.~E.,  1999, \apj, 517, 627

\bibitem[\protect\citeauthoryear{{Monaghan} \& {Lattanzio}}{{Monaghan} \&
  {Lattanzio}}{1985}]{monaghan85}
{Monaghan} J.~J.,  {Lattanzio} J.~C.,  1985, \aap, \href
  {http://adsabs.harvard.edu/abs/1985A%26A...149..135M} {149, 135}

\bibitem[\protect\citeauthoryear{{Motl}, {Hallman}, {Burns}  \&
  {Norman}}{{Motl} et~al.}{2005}]{motl05}
{Motl} P.~M.,  {Hallman} E.~J.,  {Burns} J.~O.,   {Norman} M.~L.,  2005,
  \mn@doi [\apjl] {10.1086/430144}, \href
  {http://adsabs.harvard.edu/abs/2005ApJ...623L..63M} {623, L63}

\bibitem[\protect\citeauthoryear{{Nagai}}{{Nagai}}{2006}]{nagai06}
{Nagai} D.,  2006, \mn@doi [\apj] {10.1086/506467}, \href
  {http://adsabs.harvard.edu/abs/2006ApJ...650..538N} {650, 538}

\bibitem[\protect\citeauthoryear{{Nagai}, {Kravtsov}  \& {Vikhlinin}}{{Nagai}
  et~al.}{2007}]{nagai07}
{Nagai} D.,  {Kravtsov} A.~V.,   {Vikhlinin} A.,  2007, \mn@doi [\apj]
  {10.1086/521328}, \href {http://adsabs.harvard.edu/abs/2007ApJ...668....1N}
  {668, 1}

\bibitem[\protect\citeauthoryear{{Navarro}, {Frenk}  \& {White}}{{Navarro}
  et~al.}{1997}]{navarro97}
{Navarro} J.~F.,  {Frenk} C.~S.,   {White} S.~D.~M.,  1997, \mn@doi [\apj]
  {10.1086/304888}, \href {http://adsabs.harvard.edu/abs/1997ApJ...490..493N}
  {490, 493}

\bibitem[\protect\citeauthoryear{{Nelson}, {Lau}  \& {Nagai}}{{Nelson}
  et~al.}{2014}]{nelson14}
{Nelson} K.,  {Lau} E.~T.,   {Nagai} D.,  2014, \mn@doi [\apj]
  {10.1088/0004-637X/792/1/25}, \href
  {http://adsabs.harvard.edu/abs/2014ApJ...792...25N} {792, 25}

\bibitem[\protect\citeauthoryear{{Nurgaliev} et~al.,}{{Nurgaliev}
  et~al.}{2016}]{nurgaliev16}
{Nurgaliev} D.,  et~al., 2016, preprint, \href
  {http://adsabs.harvard.edu/abs/2016arXiv160900375N} {} (\mn@eprint {arXiv}
  {1609.00375})

\bibitem[\protect\citeauthoryear{{Pfrommer}}{{Pfrommer}}{2008}]{pfrommer08}
{Pfrommer} C.,  2008, \mn@doi [\mnras] {10.1111/j.1365-2966.2008.12957.x},
  \href {http://adsabs.harvard.edu/abs/2008MNRAS.385.1242P} {385, 1242}

\bibitem[\protect\citeauthoryear{{Piffaretti}, {Arnaud}, {Pratt},
  {Pointecouteau}  \& {Melin}}{{Piffaretti} et~al.}{2011}]{piffaretti11}
{Piffaretti} R.,  {Arnaud} M.,  {Pratt} G.~W.,  {Pointecouteau} E.,   {Melin}
  J.-B.,  2011, \mn@doi [\aap] {10.1051/0004-6361/201015377}, \href
  {http://adsabs.harvard.edu/abs/2011A%26A...534A.109P} {534, A109}

\bibitem[\protect\citeauthoryear{{Plagge} et~al.,}{{Plagge}
  et~al.}{2010}]{plagge10}
{Plagge} T.,  et~al., 2010, \mn@doi [\apj] {10.1088/0004-637X/716/2/1118},
  \href {http://adsabs.harvard.edu/abs/2010ApJ...716.1118P} {716, 1118}

\bibitem[\protect\citeauthoryear{{Planck Collaboration}}{{Planck
  Collaboration}}{2006}]{planck06}
{Planck Collaboration} 2006, ArXiv:astro-ph/0604069, \href
  {http://adsabs.harvard.edu/abs/2006astro.ph..4069T} {}

\bibitem[\protect\citeauthoryear{{Planck Collaboration}}{{Planck
  Collaboration}}{2011}]{planck11-10}
{Planck Collaboration} 2011, \mn@doi [\aap] {10.1051/0004-6361/201116457},
  \href {http://adsabs.harvard.edu/abs/2011A%26A...536A..10P} {536, A10}

\bibitem[\protect\citeauthoryear{{Planck Collaboration} et~al.,}{{Planck
  Collaboration} et~al.}{2013a}]{planck12-10}
{Planck Collaboration} et~al., 2013a, \mn@doi [\aap]
  {10.1051/0004-6361/201220040}, \href
  {http://adsabs.harvard.edu/abs/2013A%26A...550A.131P} {550, A131}

\bibitem[\protect\citeauthoryear{{Planck Collaboration} et~al.,}{{Planck
  Collaboration} et~al.}{2013b}]{planck13-11}
{Planck Collaboration} et~al., 2013b, \mn@doi [\aap]
  {10.1051/0004-6361/201220941}, \href
  {http://adsabs.harvard.edu/abs/2013A%26A...557A..52P} {557, A52}

\bibitem[\protect\citeauthoryear{{Planck Collaboration} et~al.,}{{Planck
  Collaboration} et~al.}{2014}]{planck13-20}
{Planck Collaboration} et~al., 2014, \mn@doi [\aap]
  {10.1051/0004-6361/201321521}, \href
  {http://adsabs.harvard.edu/abs/2014A%26A...571A..20P} {571, A20}

\bibitem[\protect\citeauthoryear{{Planck Collaboration} et~al.,}{{Planck
  Collaboration} et~al.}{2016a}]{planck16-24}
{Planck Collaboration} et~al., 2016a, \mn@doi [\aap]
  {10.1051/0004-6361/201525833}, \href
  {http://adsabs.harvard.edu/abs/2016A%26A...594A..24P} {594, A24}

\bibitem[\protect\citeauthoryear{{Planck Collaboration} et~al.,}{{Planck
  Collaboration} et~al.}{2016b}]{planck16SZcat}
{Planck Collaboration} et~al., 2016b, \mn@doi [\aap]
  {10.1051/0004-6361/201525823}, \href
  {http://adsabs.harvard.edu/abs/2016A%26A...594A..27P} {594, A27}

\bibitem[\protect\citeauthoryear{{Rasia}, {Tormen}  \& {Moscardini}}{{Rasia}
  et~al.}{2004}]{rasia04}
{Rasia} E.,  {Tormen} G.,   {Moscardini} L.,  2004, \mn@doi [\mnras]
  {10.1111/j.1365-2966.2004.07775.x}, \href
  {http://adsabs.harvard.edu/abs/2004MNRAS.351..237R} {351, 237}

\bibitem[\protect\citeauthoryear{{Rasia} et~al.,}{{Rasia}
  et~al.}{2006}]{rasia06}
{Rasia} E.,  et~al., 2006, \mn@doi [\mnras] {10.1111/j.1365-2966.2006.10466.x},
  \href {http://adsabs.harvard.edu/abs/2006MNRAS.369.2013R} {369, 2013}

\bibitem[\protect\citeauthoryear{{Rasia} et~al.,}{{Rasia}
  et~al.}{2012}]{rasia12}
{Rasia} E.,  et~al., 2012, \mn@doi [New Journal of Physics]
  {10.1088/1367-2630/14/5/055018}, \href
  {http://adsabs.harvard.edu/abs/2012NJPh...14e5018R} {14, 055018}

\bibitem[\protect\citeauthoryear{{Reichardt} et~al.,}{{Reichardt}
  et~al.}{2013}]{reichardt13}
{Reichardt} C.~L.,  et~al., 2013, \mn@doi [\apj] {10.1088/0004-637X/763/2/127},
  \href {http://adsabs.harvard.edu/abs/2013ApJ...763..127R} {763, 127}

\bibitem[\protect\citeauthoryear{Rephaeli}{Rephaeli}{1995}]{rephaeli95}
Rephaeli Y.,  1995, \araa, 33, 541

\bibitem[\protect\citeauthoryear{{Rozo} \& {Rykoff}}{{Rozo} \&
  {Rykoff}}{2014}]{rozo14}
{Rozo} E.,  {Rykoff} E.~S.,  2014, \mn@doi [\apj] {10.1088/0004-637X/783/2/80},
  \href {http://adsabs.harvard.edu/abs/2014ApJ...783...80R} {783, 80}

\bibitem[\protect\citeauthoryear{{Rozo}, {Evrard}, {Rykoff}  \&
  {Bartlett}}{{Rozo} et~al.}{2014a}]{rozo14b}
{Rozo} E.,  {Evrard} A.~E.,  {Rykoff} E.~S.,   {Bartlett} J.~G.,  2014a,
  \mn@doi [\mnras] {10.1093/mnras/stt2160}, \href
  {http://adsabs.harvard.edu/abs/2014MNRAS.438...62R} {438, 62}

\bibitem[\protect\citeauthoryear{{Rozo}, {Bartlett}, {Evrard}  \&
  {Rykoff}}{{Rozo} et~al.}{2014b}]{rozo14a}
{Rozo} E.,  {Bartlett} J.~G.,  {Evrard} A.~E.,   {Rykoff} E.~S.,  2014b,
  \mn@doi [\mnras] {10.1093/mnras/stt2161}, \href
  {http://adsabs.harvard.edu/abs/2014MNRAS.438...78R} {438, 78}

\bibitem[\protect\citeauthoryear{{Sanderson}, {Edge}  \& {Smith}}{{Sanderson}
  et~al.}{2009}]{sanderson09}
{Sanderson} A.~J.~R.,  {Edge} A.~C.,   {Smith} G.~P.,  2009, \mn@doi [\mnras]
  {10.1111/j.1365-2966.2009.15214.x}, \href
  {http://adsabs.harvard.edu/abs/2009MNRAS.398.1698S} {398, 1698}

\bibitem[\protect\citeauthoryear{{Saro} et~al.,}{{Saro} et~al.}{2014}]{saro14}
{Saro} A.,  et~al., 2014, \mn@doi [\mnras] {10.1093/mnras/stu575}, \href
  {http://adsabs.harvard.edu/abs/2014MNRAS.440.2610S} {440, 2610}

\bibitem[\protect\citeauthoryear{{Saro} et~al.,}{{Saro} et~al.}{2015}]{saro15}
{Saro} A.,  et~al., 2015, \mn@doi [\mnras] {10.1093/mnras/stv2141}, \href
  {http://adsabs.harvard.edu/abs/2015MNRAS.454.2305S} {454, 2305}

\bibitem[\protect\citeauthoryear{{Saro} et~al.,}{{Saro} et~al.}{2016}]{saro16}
{Saro} A.,  et~al., 2016, preprint, \href
  {http://adsabs.harvard.edu/abs/2016arXiv160508770S} {} (\mn@eprint {arXiv}
  {1605.08770})

\bibitem[\protect\citeauthoryear{{Sayers} et~al.,}{{Sayers}
  et~al.}{2016}]{sayers16}
{Sayers} J.,  et~al., 2016, \mn@doi [\apj] {10.3847/0004-637X/832/1/26}, \href
  {http://adsabs.harvard.edu/abs/2016ApJ...832...26S} {832, 26}

\bibitem[\protect\citeauthoryear{{Schaye} et~al.,}{{Schaye}
  et~al.}{2010}]{schaye09}
{Schaye} J.,  et~al., 2010, \mn@doi [\mnras]
  {10.1111/j.1365-2966.2009.16029.x}, \href
  {http://adsabs.harvard.edu/abs/2010MNRAS.402.1536S} {402, 1536}

\bibitem[\protect\citeauthoryear{{Schrabback} et~al.,}{{Schrabback}
  et~al.}{2016}]{schrabback16}
{Schrabback} T.,  et~al., 2016, preprint, \href
  {http://adsabs.harvard.edu/abs/2016arXiv161103866S} {} (\mn@eprint {arXiv}
  {1611.03866})

\bibitem[\protect\citeauthoryear{{Sehgal} et~al.,}{{Sehgal}
  et~al.}{2011}]{sehgal11}
{Sehgal} N.,  et~al., 2011, \mn@doi [\apj] {10.1088/0004-637X/732/1/44}, \href
  {http://adsabs.harvard.edu/abs/2011ApJ...732...44S} {732, 44}

\bibitem[\protect\citeauthoryear{{Sehgal} et~al.,}{{Sehgal}
  et~al.}{2013}]{sehgal13}
{Sehgal} N.,  et~al., 2013, \mn@doi [\apj] {10.1088/0004-637X/767/1/38}, \href
  {http://adsabs.harvard.edu/abs/2013ApJ...767...38S} {767, 38}

\bibitem[\protect\citeauthoryear{{Shaw}, {Holder}  \& {Bode}}{{Shaw}
  et~al.}{2008}]{shaw08}
{Shaw} L.~D.,  {Holder} G.~P.,   {Bode} P.,  2008, \mn@doi [\apj]
  {10.1086/589849}, \href {http://adsabs.harvard.edu/abs/2008ApJ...686..206S}
  {686, 206}

\bibitem[\protect\citeauthoryear{{Shi} \& {Komatsu}}{{Shi} \&
  {Komatsu}}{2014}]{shi14}
{Shi} X.,  {Komatsu} E.,  2014, \mn@doi [\mnras] {10.1093/mnras/stu858}, \href
  {http://adsabs.harvard.edu/abs/2014MNRAS.442..521S} {442, 521}

\bibitem[\protect\citeauthoryear{{Shi}, {Komatsu}, {Nelson}  \& {Nagai}}{{Shi}
  et~al.}{2015}]{shi15}
{Shi} X.,  {Komatsu} E.,  {Nelson} K.,   {Nagai} D.,  2015, \mn@doi [\mnras]
  {10.1093/mnras/stv036}, \href
  {http://adsabs.harvard.edu/abs/2015MNRAS.448.1020S} {448, 1020}

\bibitem[\protect\citeauthoryear{{Short} \& {Thomas}}{{Short} \&
  {Thomas}}{2009}]{short09}
{Short} C.~J.,  {Thomas} P.~A.,  2009, \mn@doi [\apj]
  {10.1088/0004-637X/704/2/915}, \href
  {http://adsabs.harvard.edu/abs/2009ApJ...704..915S} {704, 915}

\bibitem[\protect\citeauthoryear{{Sijacki}, {Pfrommer}, {Springel}  \&
  {En{\ss}lin}}{{Sijacki} et~al.}{2008}]{sijacki08}
{Sijacki} D.,  {Pfrommer} C.,  {Springel} V.,   {En{\ss}lin} T.~A.,  2008,
  \mn@doi [\mnras] {10.1111/j.1365-2966.2008.13310.x}, \href
  {http://adsabs.harvard.edu/abs/2008MNRAS.387.1403S} {387, 1403}

\bibitem[\protect\citeauthoryear{{Song} et~al.,}{{Song} et~al.}{2012}]{song12b}
{Song} J.,  et~al., 2012, \mn@doi [\apj] {10.1088/0004-637X/761/1/22}, \href
  {http://adsabs.harvard.edu/abs/2012ApJ...761...22S} {761, 22}

\bibitem[\protect\citeauthoryear{{Springel}}{{Springel}}{2005}]{springel05g}
{Springel} V.,  2005, \mn@doi [\mnras] {10.1111/j.1365-2966.2005.09655.x},
  \href {http://adsabs.harvard.edu/abs/2005MNRAS.364.1105S} {364, 1105}

\bibitem[\protect\citeauthoryear{{Springel} \& {Hernquist}}{{Springel} \&
  {Hernquist}}{2002}]{springel02}
{Springel} V.,  {Hernquist} L.,  2002, \mn@doi [\mnras]
  {10.1046/j.1365-8711.2002.05445.x}, \href
  {http://adsabs.harvard.edu/abs/2002MNRAS.333..649S} {333, 649}

\bibitem[\protect\citeauthoryear{{Springel} \& {Hernquist}}{{Springel} \&
  {Hernquist}}{2003}]{springel03}
{Springel} V.,  {Hernquist} L.,  2003, \mn@doi [\mnras]
  {10.1046/j.1365-8711.2003.06206.x}, \href
  {http://adsabs.harvard.edu/abs/2003MNRAS.339..289S} {339, 289}

\bibitem[\protect\citeauthoryear{{Springel}, {White}, {Tormen}  \&
  {Kauffmann}}{{Springel} et~al.}{2001}]{springel01}
{Springel} V.,  {White} S.~D.~M.,  {Tormen} G.,   {Kauffmann} G.,  2001,
  \mn@doi [\mnras] {10.1046/j.1365-8711.2001.04912.x}, \href
  {http://adsabs.harvard.edu/abs/2001MNRAS.328..726S} {328, 726}

\bibitem[\protect\citeauthoryear{{Staniszewski} et~al.,}{{Staniszewski}
  et~al.}{2009}]{staniszewski09}
{Staniszewski} Z.,  et~al., 2009, \mn@doi [\apj] {10.1088/0004-637X/701/1/32},
  \href {http://adsabs.harvard.edu/abs/2009ApJ...701...32S} {701, 32}

\bibitem[\protect\citeauthoryear{{Stott} et~al.,}{{Stott}
  et~al.}{2012}]{stott12}
{Stott} J.~P.,  et~al., 2012, \mn@doi [\mnras]
  {10.1111/j.1365-2966.2012.20764.x}, \href
  {http://adsabs.harvard.edu/abs/2012MNRAS.422.2213S} {422, 2213}

\bibitem[\protect\citeauthoryear{{Sun}, {Sehgal}, {Voit}, {Donahue}, {Jones},
  {Forman}, {Vikhlinin}  \& {Sarazin}}{{Sun} et~al.}{2011}]{sun11}
{Sun} M.,  {Sehgal} N.,  {Voit} G.~M.,  {Donahue} M.,  {Jones} C.,  {Forman}
  W.,  {Vikhlinin} A.,   {Sarazin} C.,  2011, \mn@doi [\apjl]
  {10.1088/2041-8205/727/2/L49}, \href
  {http://adsabs.harvard.edu/abs/2011ApJ...727L..49S} {727, L49+}

\bibitem[\protect\citeauthoryear{{Sunyaev} \& {Zel'dovich}}{{Sunyaev} \&
  {Zel'dovich}}{1970}]{sunyaev70}
{Sunyaev} R.~A.,  {Zel'dovich} Y.~B.,  1970, Comments on Astrophysics and Space
  Physics, \href {http://adsabs.harvard.edu/abs/1970CoASP...2...66S} {2, 66}

\bibitem[\protect\citeauthoryear{{Sunyaev} \& {Zel'dovich}}{{Sunyaev} \&
  {Zel'dovich}}{1972}]{sunyaev72}
{Sunyaev} R.~A.,  {Zel'dovich} Y.~B.,  1972, Comments on Astrophysics and Space
  Physics, \href
  {http://adsabs.harvard.edu/cgi-bin/nph-bib_query?bibcode=1972CoASP...4..173S&amp;db_key=AST}
  {4, 173}

\bibitem[\protect\citeauthoryear{{Teklu}, {Remus}, {Dolag}, {Beck}, {Burkert},
  {Schmidt}, {Schulze}  \& {Steinborn}}{{Teklu} et~al.}{2015}]{teklu15}
{Teklu} A.~F.,  {Remus} R.-S.,  {Dolag} K.,  {Beck} A.~M.,  {Burkert} A.,
  {Schmidt} A.~S.,  {Schulze} F.,   {Steinborn} L.~K.,  2015, \mn@doi [\apj]
  {10.1088/0004-637X/812/1/29}, \href
  {http://adsabs.harvard.edu/abs/2015ApJ...812...29T} {812, 29}

\bibitem[\protect\citeauthoryear{{Tescari}, {Viel}, {Tornatore}  \&
  {Borgani}}{{Tescari} et~al.}{2009}]{tescari09}
{Tescari} E.,  {Viel} M.,  {Tornatore} L.,   {Borgani} S.,  2009, \mn@doi
  [\mnras] {10.1111/j.1365-2966.2009.14943.x}, \href
  {http://adsabs.harvard.edu/abs/2009MNRAS.397..411T} {397, 411}

\bibitem[\protect\citeauthoryear{{Tinker}, {Kravtsov}, {Klypin}, {Abazajian},
  {Warren}, {Yepes}, {Gottl{\"o}ber}  \& {Holz}}{{Tinker}
  et~al.}{2008}]{tinker08}
{Tinker} J.,  {Kravtsov} A.~V.,  {Klypin} A.,  {Abazajian} K.,  {Warren} M.,
  {Yepes} G.,  {Gottl{\"o}ber} S.,   {Holz} D.~E.,  2008, \mn@doi [\apj]
  {10.1086/591439}, \href {http://adsabs.harvard.edu/abs/2008ApJ...688..709T}
  {688, 709}

\bibitem[\protect\citeauthoryear{{Tornatore}, {Borgani}, {Springel},
  {Matteucci}, {Menci}  \& {Murante}}{{Tornatore} et~al.}{2003}]{tornatore07a}
{Tornatore} L.,  {Borgani} S.,  {Springel} V.,  {Matteucci} F.,  {Menci} N.,
  {Murante} G.,  2003, \mn@doi [\mnras] {10.1046/j.1365-8711.2003.06631.x},
  \href {http://adsabs.harvard.edu/abs/2003MNRAS.342.1025T} {342, 1025}

\bibitem[\protect\citeauthoryear{{Tornatore}, {Ferrara}  \&
  {Schneider}}{{Tornatore} et~al.}{2007a}]{tornatore07c}
{Tornatore} L.,  {Ferrara} A.,   {Schneider} R.,  2007a, \mn@doi [\mnras]
  {10.1111/j.1365-2966.2007.12215.x}, \href
  {http://adsabs.harvard.edu/abs/2007MNRAS.382..945T} {382, 945}

\bibitem[\protect\citeauthoryear{{Tornatore}, {Borgani}, {Dolag}  \&
  {Matteucci}}{{Tornatore} et~al.}{2007b}]{tornatore07b}
{Tornatore} L.,  {Borgani} S.,  {Dolag} K.,   {Matteucci} F.,  2007b, \mn@doi
  [\mnras] {10.1111/j.1365-2966.2007.12070.x}, \href
  {http://adsabs.harvard.edu/abs/2007MNRAS.382.1050T} {382, 1050}

\bibitem[\protect\citeauthoryear{{Vanderlinde} et~al.,}{{Vanderlinde}
  et~al.}{2010}]{vanderlinde10}
{Vanderlinde} K.,  et~al., 2010, \mn@doi [\apj] {10.1088/0004-637X/722/2/1180},
  \href {http://adsabs.harvard.edu/abs/2010ApJ...722.1180V} {722, 1180}

\bibitem[\protect\citeauthoryear{{White}, {Hernquist}  \& {Springel}}{{White}
  et~al.}{2002}]{white02}
{White} M.,  {Hernquist} L.,   {Springel} V.,  2002, \mn@doi [\apj]
  {10.1086/342756}, \href
  {http://adsabs.harvard.edu/cgi-bin/nph-bib_query?bibcode=2002ApJ...579...16W&db_key=AST}
  {579, 16}

\bibitem[\protect\citeauthoryear{{Wiersma}, {Schaye}  \& {Smith}}{{Wiersma}
  et~al.}{2009}]{wiersma09}
{Wiersma} R.~P.~C.,  {Schaye} J.,   {Smith} B.~D.,  2009, \mn@doi [\mnras]
  {10.1111/j.1365-2966.2008.14191.x}, \href
  {http://adsabs.harvard.edu/abs/2009MNRAS.393...99W} {393, 99}

\bibitem[\protect\citeauthoryear{de Haan et~al.,}{de~Haan
  et~al.}{2016}]{dehaan16}
de Haan T.,  et~al., 2016, The Astrophysical Journal, 832, 95

\bibitem[\protect\citeauthoryear{{von der Linden} et~al.,}{{von der Linden}
  et~al.}{2014}]{vonderlinden14}
{von der Linden} A.,  et~al., 2014, \mn@doi [\mnras] {10.1093/mnras/stu1423},
  \href {http://adsabs.harvard.edu/abs/2014MNRAS.443.1973V} {443, 1973}

\makeatother
\end{thebibliography}

\end{document}